\documentclass[reprint,amsmath,amssymb, aps,prl,floatfix,superscriptaddress]{revtex4-2}
\usepackage{graphicx}
\usepackage{xcolor}
\usepackage{amsmath}

\newcommand{\env}[1]{\texttt{#1}}

\usepackage{docs}
\usepackage{nameref}
\usepackage[colorlinks=true,linkcolor=blue]{hyperref}

\begin{document}


\title{Emergence of run-and-tumble-like swimming in self-propelling artificial swimmers in soft microchannels}
\author{Smita S. Sontakke}%
\affiliation{
 Department of Mechanical and Aerospace Engineering,\\
Indian Institute of Technology Hyderabad,
Kandi, Sangareddy 502285, India }
\author{Aneesha Kajampady}%
\affiliation{
 Department of Mechanical and Aerospace Engineering,\\
Indian Institute of Technology Hyderabad,
Kandi, Sangareddy 502285, India }
\author{Mohd Suhail Rizvi}%
\email{suhailr@bme.iith.ac.in }
\affiliation{
 Department of Biomedical Engineering,\\
Indian Institute of Technology Hyderabad,
Kandi, Sangareddy 502285, India }
\author{Ranabir Dey}%
\email{ranabir@mae.iith.ac.in}
\affiliation{ 
Department of Mechanical and Aerospace Engineering,\\
Indian Institute of Technology Hyderabad,
Kandi, Sangareddy 502285, India }
%

\begin{abstract}
Biological microswimmers often encounter deformable boundaries in physiological conditions; for instance, the viscoelastic walls of reproductive tract during migration of spermatozoa, or host tissue during early bacterial biofilm formation. 
However, the combined influence of elastic and hydrodynamic cues on microswimmer dynamics is poorly understood.
Here, we experimentally investigate how the softness of microchannel walls affects the swimming characteristics of self-propelling microswimmers, using autophoretic active droplets as a model system.    
Remarkably, in a soft microchannel, a self-propelling droplet exhibits a run-and-tumble-like motility characterized by abrupt reorientations in the swimming direction, which are accompanied by local reduction and subsequent increase in the swimming speed. 
Such emergent swimming dynamics in response to increasing softness of microchannels have been previously unobserved for synthetic microswimmers.
Using 3D boundary integral simulations and fluorescence microscopy experiments, we show that the coupling between the elastohydrodynamic interactions and the chemo-hydrodynamics, inherent in the self-propulsion mechanism, in a soft narrow confinement results in alterations in the swimming characteristics.
We envisage that such adaptation of autophoretic microswimmers to changes in the softness of microchannel walls will pave the way for novel methods for tuning active agents in complex environment solely by exploiting the elasticity of confining walls. 

\end{abstract}
\maketitle
\section{Introduction}
Formation of bacterial biofilms on soft living tissues culminates in an overwhelming number of critical diseases, e.g. bacterial prostatitis caused by E. coli, cystic fibrosis pneumonia caused by Pseudomonas aeruginosa, and infections of the gastrointestinal tract \cite{costerton1999bacterial,rossi2018sa}.
An important step in the commencement of such biofilm formation is the initial reversible attachment of a free-swimming bacterium onto a soft surface using adhesins such as flagellum or pilli \cite{petrova2012sticky, krsmanovic2021hydrodynamics}. 
The attachment of bacteria to a surface is dependent on the intervening hydrodynamic interactions and surface characteristics involved, such as material stiffness and topography \cite{krsmanovic2021hydrodynamics, dufrene2020mechanomicrobiology}.

Only recently, the effects of the softness of an underlying substrate \cite{fei2020nonuniform} and of a soft elastic confinement \cite{fortune2022biofilm} on the 3D morphology of a biofilm have been studied. 
However, interestingly, the relationship between the initial attachment of swimming bacteria and substrate softness (stiffness) has remained ambiguous, despite its significance in the initiation of biofilms. 
On the one hand, some research works demonstrated that the adhesion density of swimming bacteria, such as E. coli, increased with increasing softness (reducing stiffness; Young's modulus in the range of $\sim \mathcal{O}(0.01)-\mathcal{O}(1)$ MPa) of a confining polymeric (polydimethylsiloxane, PDMS) substrate, suggesting that softer substrates favour the planktonic to sessile transition of bacteria \cite{saha2013influence, song2014stiffness, song2017bacteria}.
While on the other, some research works showed that the adhesion density of similar bacteria, such as E. coli, reduced with increasing softness of a polymeric substrate over a comparable range of Young's modulus, implying the propensity for adhesion to stiffer substrates \cite{lichter2008substrata, guegan2014alteration, kolewe2015fewer, peng2019three}.
Despite such apparent contradiction, the general consensus is that the near-wall swimming characteristics of microorganisms, like tumbling/flicking frequency controlled by bacterial flagellar motor, are altered by the hydrodynamic interactions with a soft substrate \cite{song2017bacteria, peng2019three}.
However, the physical aspects of the interaction between a swimming microorganism and a soft, deformable substrate remains poorly understood. 
A clear understanding of the interaction is difficult because of the multidisciplinary aspect, at the crossroads of biology, chemistry, fluid mechanics, and solid mechanics. 

Over the years, self-propelling synthetic microswimmers, specifically catalytic Janus particles \cite{paxton2004catalytic, howse2007self, zottl2016_emergent, campbell2019experimental, liebchen2021_interactions, zottl2023_modeling}  and active droplets \cite{toyota2009_selfpropelled, Swarmingbehavior2011, peddireddy2012solubilization, herminghaus2014interfacial, izri2014self, Swimmingdroplets2016, birrer2022_we, dwivedi2022_selfpropelled,michelin2023_selfpropulsion}, have emerged as model systems for probing the chemo-hydrodynamic interactions of microswimmers in complex environment. 
Catalytic Janus particles are driven by phoretic mechanisms \cite{moran2017phoretic}, like self-diffusiophoresis \cite{self-diffusiophoresis}, self-electrophoresis \cite{pumera2010electrochemically}, or self-thermophoresis \cite{jiang2010active, yu2019phototaxis}.
These mechanisms are triggered by chemical reactions along the particle's boundary, which is engineered with fore-aft asymmetry. 
In contrast, active droplets are isotropic, autophoretic microswimmers that self-propel due to spontaneous creation of an interfacial tension gradient along the droplet interface by symmetry-breaking of surfactant concentration \cite{peddireddy2012solubilization, herminghaus2014interfacial, izri2014self, Swimmingdroplets2016, michelin2023_selfpropulsion, michelin2013spontaneous}. 
Fascinatingly, like biological microswimmers, both types of synthetic microswimmers can also alter their swimming characteristics in response to external cues such as changes in chemical concentration \cite{sundararajan2008catalytic, jin2017chemotaxis, hokmabad2022chemotactic}, changes in viscosity of the swimming medium \cite{hokmabad2021emergence}, external flows \cite{uspal2015rheotaxis, katuri2018cross, dwivedi2021rheotaxis, dey2022oscillatory}, and external fields (gravity, electric) \cite{castonguay2023gravitational, buness2024electrotaxis}.
These investigations have helped to unveil the possible chemo-hydrodynamic mechanisms hidden in complex biophysical processes like chemotaxis \cite{berg1972chemotaxis}, rheotaxis \cite{mathijssen2019oscillatory}, gravitaxis \cite{sengupta2023planktonic}, and electrotaxis \cite{kim2013galvanotactic}.
However, even for such synthetic model microswimmers, the understanding of near-wall swimming dynamics has remained confined to rigid walls \cite{das2015boundaries, de2019flow, ketzetzi2020slip, ketzetzi2020diffusion}, curved boundaries \cite{jin2019fine}, and in narrow confinements \cite{liu2016bimetallic, de2021swimming, guchhait2025flow}.
The swimming characteristics of self-propelling artificial microswimmers near soft, deformable walls or in soft, narrow confinements have never been studied so far.

It has been theoretically argued that deformation of an interface, such as that of an elastic gel, due to far-field hydrodynamic interactions triggered by the reciprocal motion of a particle in effective microscale swimming, thus breaking the constraints of the Scallop theorem \cite{trouilloud2008soft}. 
The connection between the deformation of an elastic interface and the motility of a microswimmer was further investigated using 3D numerical simulations of a model dipolar microswimmer in an elastic capillary \cite{ledesma2013enhanced}.
The deformation of the elastic capillary surface due to hydrodynamic interaction enhances the speed and efficiency of a model dipolar swimmer, when the propulsion time scale is smaller or comparable to the boundary relaxation time scale \cite{ledesma2013enhanced}.
In contrast, the average swimming speed of an amoeboid (shape changing) microswimmer in a flexible microchannel was numerically shown to be smaller than that in a rigid microchannel \cite{dalal2020amoeboid}.
Efforts were also made to theoretically study the swimming dynamics of a microswimmer near a 2D deformable interface, having both shear and bending resistances, using far-field hydrodynamic model \cite{daddi2019frequency} and slender body theory \cite{nambiar2022hydrodynamics}. 
Only recently, the coupling between the swimming dynamics of a microswimmer and the deformation of an elastic boundary was studied using the infinite 2D Taylor's sheet model near a Winkler's mattress \cite{jha2025taylor}.
For the model microswimmer, the swimming velocity of transverse (longitudinal) waves decreased (increased) with increasing softness of the confining boundary \cite{jha2025taylor}.
Hence, the few existing theoretical studies are restricted by the idealization of the microswimmer (e.g. far-field approximations, Taylor's sheet) and/or by the consideration of 2D interfaces/membranes.
The coupling between the swimming dynamics of a finite-sized model microswimmer near a soft, deformable wall, or in a soft microconfinement, and the elastohydrodynamic interactions has remained overwhelmingly understudied. 

In this study, we report a quantitative experimental investigation into the swimming dynamics of an active droplet, which is a finite-sized, autophoretic microswimmer, in microchannels with soft, deformable walls. 
Remarkably, in a soft microchannel, the active droplet autonomously alters its motility to a run-and-tumble-like swimming, in contrast to a steady, unidirectional swimming along a wall in a rigid microchannel. 
The emergent swimming dynamics in the soft microchannel is characterized by intermittent reorientations in the swimming direction (`tumbles'), preceded and followed by local variations in the swimming speed. 
We use boundary-integral method-based 3D numerical simulations to reveal the elastohydrodynamic interaction between the finite-sized model microswimmer and the soft, deformable walls of a microconfinement, leading to the slowing down of the microswimmer.
Thereafter, using precise microscopy experiments, we argue that the advection-dominated growth of the self-generated chemical exhaust over the posterior periphery of such a slow-moving microswimmer results in further reduction in swimming speed, and eventual reorientation.
We think that this is the first work which shows that the intimate coupling between elastohydrodynamic and chemo-hydrodynamic interactions can result in altered motility states for an isotropic, autophoretic microswimmer in a soft microchannel.
This new physical understanding can have two important consequences- one, it can be utilized for conceptuatlizing state-of-the-art in vivo applications like targeted cargo delivery where autonomous navigation in soft, narrow confinements will be the norm; and two, it will shed new light on the role of elastohydrodynamic interactions in altering the swimming dynamics of biological microswimmers near soft deformable boundaries. 

\section{Experimental setup and methodology}
\subsection*{Fabrication of soft microchannels}
We fabricate polydimethylsiloxane (PDMS) (SYLGARD 184; DOW Chemicals) based elastomeric microchannels of increasing softness (reducing Young's modulus E) using standard softlithography technique with prepolymer having increasing base polymer to cross-linker ratio-- 10:1, 25:1, and 40:1. The prepolymer mixture is poured onto a 3D-printed (micro-SLA; Boston Micro Fabrication) mold, which contains the negative replica of the microchannel. 
The 10:1 prepolymer is cured at 75 $^\circ$C for 3 hours in a hot air oven, while the 25:1 and the 40:1 prepolymers are cured at 75 $^\circ$C for 17 hours.
The rheology of cured PDMS elastomers is tested separately using an Anton Paar Rheometer (see Supplementary Material, SM-Fig. \ref{fgr:SM2}(a) \cite{sontakke2025_supplemental}). 
For the cured PDMS polymer, $E$ reduces from $360$ kPa to $35.2$ kPa as the base polymer-crosslinker ratio increases from 10:1 to 40:1.
The cured stamps are bonded to cover slips coated with cured PDMS films having identical $E$ as the stamp.

The rectangular micro-channels are $2w = 160$ $\mu$m in width, $h = 100$ $\mu$m in height, and $1$ cm in length (Fig. \ref{fig:1}(a); also see SM-Fig. \ref{fgr:SM2}(b) \cite{sontakke2025_supplemental}). 

\subsection*{Generation of active droplets}
The active droplets are generated using a microfluidic flow-focusing device in which two jets of continuous phase shear the central jet of the dispersed phase (SM-Figs. \ref{fgr:SM1}(a) and (b) \cite{sontakke2025_supplemental}). 
We use 0.1 wt$\%$ aqueous surfactant (Trimethyl(tetradecyl)ammonium bromide, TTAB; Sigma-Aldrich) solution, having surfactant concentration lower than the critical miceller concentration (CMC=0.13 wt$\%$), as the continuous phase, and CB-15 ((S)-4-Cyano-4$'$-(2-methyl butyl) biphenyl; Tokiyo Chemical Industry) as the dispersed phase.
The flow rates of the CB-15 and 0.1 wt$\%$ aqueous TTAB solution are judiciously tuned using syringe pumps (F101 and F100X; Chemyx) to obtain monodispersed CB15 droplets. 
The generated droplet sizes are in the range of $R_d=29.2 \pm 0.8$ $\mu$m. 
We make these oil droplets active, i.e. self-propelling, by further mixing them in 7.5 wt$\%$ (greater than CMC) aqueous TTAB solution in 1:5 ratio during experiments.

These active droplets self-propel due to micellar solubilization \cite{peddireddy2012solubilization, herminghaus2014interfacial,  Swimmingdroplets2016, michelin2023_selfpropulsion}.
When a CB15 droplet, as generated above, is dispersed in a 7.5 wt$\%$ aqueous TTAB solution, its interface is initially uniformly covered with surfactant monomers.
During the solubilization process, the empty micelles in the solution take up oil, along with surfactant monomers,  from the interfacial region of the droplet and transform into filled micelles.
Any advective perturbation in the system disturbs the uniform distribution of filled micelles, and hence, the uniform surfactant coverage, creating a nonuniform surfactant concentration along the droplet interface.
The resulting surface tension gradient induces Marangoni stress, which drives a net flow of the aqueous solution near the droplet's interfacial region towards the area of higher interfacial tension, i.e., towards the region where filled micelle (surfactant monomer) concentration is higher (lower). 
To satisfy the condition of a non-inertial system at the microscale, the droplet self-propels in the direction opposite to the flow caused by the surface tension gradient, i.e. towards a higher concentration of empty micelles.
The active droplets swim in a quasi-ballistic manner, and leave behind a wake of filled micelles \cite{hokmabad2021emergence, hokmabad2022chemotactic}. 

\subsection*{Bright field microscopy}
We characterize the trajectories of the self-propelling active droplets in microchannels by tracking the droplet centroid position over time using bright field microscopy with an inverted Nikon ECLIPSE Ti microscope at 20$\times$ magnification. 
The image sequences are recorded at 25 frames per second (fps) with a CMOS monochrome digital camera (IDS Imaging) having $1920 \times 1200$ pixels in the $1/1.2^{\prime\prime}$ CMOS sensor. 
The temporal evolution in the centroid position is evaluated using an in-house image processing routine (in MATLAB R2020a). 
Subsequently, from these data sets, we calculate the instantaneous swimming velocity vector $\vec{v}_d=v_x \hat{i}+ v_y \hat{j}$, using the central difference scheme, and the instantaneous swimming speed ${v}_d=|\vec{v}_d|$.
Furthermore, we also evaluate the swimming orientation in the laboratory reference frame $\hat{e}=\cos \psi \; \hat{i}+\sin \psi \; \hat{j}$ (Figs. ~\ref{fig:1}(b) and (c)), where $\psi= \tan ^{-1}(v_{y}/v_{x})$.  
The minimum translation velocity that can be detected using a $20\times$ objective (with a spatial resolution of $0.2802$ $\mu$m/pixel) and a frame capture rate of 25 fps is 3.5 $\mu$m/s.

\subsection*{\texorpdfstring{Fluorescence microscopy and micro-Particle Image Velocimetry ($\mu$-PIV)}{Fluorescence microscopy and micro-Particle Image Velocimetry (μ-PIV)}}

The velocity field generated by the droplet microswimmers is characterized using high-resolution, micro-particle image velocimetry ($\mu$-PIV) analysis. We use 500 nm carboxylate-modified fluorescent microspheres (Thermo Fisher Scientific) with excitation/emission wavelengths of 505/515 nm to seed the swimming medium.
We mix these fluorescent particles with the $7.5\%$ aqueous surfactant solution in $1:20$ volume ratio, and the solution is then sonicated for $10$ minutes to ensure proper mixing without particle coagulation. 
The active droplets in the $0.1\%$ aqueous surfactant solution are added to the sonicated $7.5\%$ surfactant solution containing tracer particles in the ratio of $1:5$. 

To visualize the flow field generated by the droplet microswimmers in the microchannels, we use the inverted Nikon ECLIPSE Ti microscope.
The microscope is fitted with a high-pressure illumination module containing a $100$ W mercury lamp (Nikon, Model$:$ LH-M100C-1). 
To get the desired excitation (emission) wavelength, a dual-band filter cube with two peak excitation and emission wavelengths- $487/562$ nm (excitation) and $523/630$ nm (emission) (Chroma Technology, Model$:$ 59009 - ET - FITC/CY3) is integrated with the microscope (Fig. ~\ref{fig:1}(a)). 
We use an open-source platform based on MATLAB- PIVlab \cite{stamhuis2014pivlab, Thielicke_2021}, for $\mu$-PIV analysis.

We extract the velocity vector field data sets after the analysis and post-process the data to plot the streamlines and velocity contours using our in-house MATLAB subroutine.
For details of the $\mu$-PIV analysis and the associated uncertainty analysis, see SM- sec. 1.3 \cite{sontakke2025_supplemental}. 
\nocite {schneider2012nih,Flow_Trace,Flow_Trace_2,guchhait2025flow, Thielicke_2021,tropea2007springer,raffel2018particle,chandrala2016unsteady,phdthesis,gui2002correlation}

\subsection*{Quantitative visualization of the chemical trail}
To characterize the concentration of filled micelles self-generated by the active droplets \cite{hokmabad2021emergence, hokmabad2022chemotactic}  self-propelling in the microchannels, we perform independent experiments under identical conditions. 
In this case, the CB15 oil is tagged with the fluorescent Nile Red dye (Invitrogen™) (excitation/emmission: $550/640$ nm), and the droplets of the same are generated using the aforementioned droplet generation technique.
Like before, these droplets are mixed with 7.5 wt$\%$ aqueous TTAB solution to make them active, and then injected into the microchannels.
We use the same inverted fluorescence microscopy setup equipped with a dual-band filter cube to visualize the filled micelle distribution around the swimming droplets.
The filled micelles contain Nile Red-tagged oil molecules and emit fluorescence.
The fluorescence image sequence is captured at 25 fps by fixing an exposure that does not over-saturate the pixels around the swimming droplet.   

We post-process the fluorescence signal intensity around the droplet interface using an independent, in-house MATLAB code. 
The fluorescence signal intensity is extracted at a distance of 22 $\mu$m away from the droplet circumference for each image.
The pixel-intensity ($I$) is extracted along the droplet circumference following a clockwise sweep starting from the positive X-direction ($\theta=0$) (see SM-Fig. \ref{fgr: SM3} \cite{sontakke2025_supplemental}).  
We evaluate the temporal variation of $I(\theta)$ $(\theta \in [0, 2 \pi])$ over time to generate the kymographs (see SM-Fig. \ref{fgr: SM3} \cite{sontakke2025_supplemental}).
The intensity values are normalized by the maximum intensity value ($I_{max}$) observed across the image sequence.
Furthermore, from these data, we quantify the temporal evolution of the fractional coverage of the microswimmer's periphery by the filled micelle wake as $\bar{\lambda}(t)=\Delta \varphi(t)/2 \pi$.
Here $\Delta \varphi(t)$ is the angular measure of the part of the circumference covered by the filled micelle wake (see SM- Fig. \ref{fgr: SM3} \cite{sontakke2025_supplemental}).

\section{Model}
The modeling framework considers a spherical squirmer of radius $R_d$ swimming in a viscous, incompressible fluid confined between two elastic half-spaces separated by distance $h$ (Figs. \ref{fig:1}(a) and (b); also see SM- Fig.  \ref{fgr: SMnum} \cite{sontakke2025_supplemental}). 
Given the microscopic nature of the system, the fluid flow is considered to be inertialess (Re $\sim$ 0), and can be described by the Stokes equations 
\begin{equation}
-\nabla p + \mu \nabla^2 \mathbf{u} = 0, \qquad \nabla \cdot \mathbf{u} = 0, \label{eq:stokes}
\end{equation}
where $p$ and $\mathbf{u}$ are pressure and velocity field, respectively, and $\mu$ is the dynamic viscosity of the fluid.
The surface activity of the squirmer drives flow in the surrounding medium, leading to hydrodynamic interactions with the elastic boundaries. 
These interactions, in turn, induce deformation in the bounding half-spaces.


\subsection*{Squirmer}
The squirmer propels itself via a prescribed tangential slip velocity on its surface, modeled using the classical axisymmetric squirmer formulation. 
In the absence of any perturbations in the micellar field (we will relax this assumption later), without loss of generality, we can assume that the swimmer propulsion direction remains in $x-z$ plane and we can specify the squirmer’s position in terms of its perpendicular distance in $z$-direction from one of the walls, and its propulsion direction in terms of the $\phi$ relative to the $x$-axis (see SM- Fig. \ref{fgr: SMnum} \cite{sontakke2025_supplemental}).  
This minimal parametrization, wall distance and orientation angle, captures the essential hydrodynamic interactions in the system while leveraging the axisymmetry of the swimmer and the translational invariance along the direction parallel to the walls.
The slip velocity $\mathbf{u}_s$ encodes the propulsive mechanism of the squirmer. 
For an axisymmetric squirmer, this is typically prescribed as 
\begin{equation}
\mathbf{u}_s(\theta) = B_1 \sin\theta\, \hat{\boldsymbol{\theta}} + B_2 \sin\theta \cos\theta\, \hat{\boldsymbol{\theta}},
\end{equation}
where $\theta$ is the angle relative to the swimmer's propulsion direction, $\hat{\boldsymbol{\theta}}$ is the unit vector in the azimuthal direction on the surface, and $B_1$, $B_2$ are scalar coefficients. 
The ratio $\beta = B_2/B_1$ characterizes the stresslet strength and hence the nature of the squirmer being a puller ($\beta > 0$) or a pusher ($\beta < 0$). 

\subsection*{Elastic boundaries}
The confining walls are treated as deformable elastic boundaries, and the domain beyond each wall is considered to be linearly elastic half-space, capturing the influence of substrate compliance on the fluid–structure interaction. 
The flow induced by the squirmer generates a pressure distribution on the walls. 
This hydrodynamic pressure serves as normal traction acting on the elastic boundaries. 
We consider the relaxation time of the elastic half space to be much smaller than the characteristic timescale of the squirmer dynamics and, therefore, the its deformation can be described by 
\begin{equation}
    \nabla \cdot \boldsymbol{\sigma} = 0
\end{equation}
where $\boldsymbol{\sigma}$ is the elastic stress and it is linked to the deformation field by 
\begin{equation}
    \boldsymbol{\sigma} = \mathbb{C} \boldsymbol{\varepsilon}
\end{equation}
where, for linear elastic solids, $\mathbb{C}$ can be described by two independent parameters - Poisson's ratio $\nu$, and elasticity coefficient E. 

To compute the resulting deformation due to squirmer activity, we use solution of the classic Boussinesq problem in elasticity \cite{johnson1987contact}. 
The Boussinesq problem links the deformation of the elastic half-space to the localized pressure applied on the boundary. 
This coupling of hydrodynamic pressure with substrate deformation 
provides a computationally efficient way 
to include boundary deformations. 

\subsection*{Numerical simulation}
We solve the fluid-structure interaction iteratively by alternating between the Stokes flow and elastic deformation equations.
We solve the fluid flow problem using boundary integral method with regularized Stokeslet \cite{cortez2001method}.
For the elastic domain we use the solution of classical Boussinesq problem to estimate deformation field. 
The interaction between the two domains is described in terms of the interfacial conditions. 
For fluid flow, the deformed elastic domain acts as the no slip boundary and for elastic medium the pressure exerted by the squirmer activity acts as the traction boundary condition. 
See Supplementary Information for the details of implementation. 

\section{Results}
We represent the increasing stiffness of the microchannel walls by the non-dimensional elastoviscous number \cite{trouilloud2008soft}, $\mathcal{P}=E R_{d} \epsilon^2/\mu v_{0}$. 
With $E$ decreasing from 360 kPa to 35.2 kPa, $\mathcal{P}$ reduces from 5 to 0.6.
Here, $E$ is the Young's modulus, $R_d=29.2 \pm 0.8$ $\mu$m is the average droplet microswimmer radius, $v_0=21.3$ $\mu $m/s is the average swimming speed $(\langle v_d \rangle)$ of the droplet microswimmer in the microchannel with the largest value of $E$ (i.e. $v_0=\langle v_d \rangle_{\mathcal{P}=5}$), and $\epsilon \sim 10^{-4}$ is a small non-dimensional quantity. 
The $\epsilon^2$ factor is there to keep the range for $\mathcal{P} \sim \mathcal{O}(0.1) - \mathcal{O}(1)$ for the ease of presentation.
We non-dimensionalize spatial variables by $R_d$, velocity by $v_0$, and time by the advective time scale $R_d/v_0$.

\subsection*{Emergence of run-and-tumble-like swimming in soft microchannels}
\begin{figure*}[ht]
 \centering
 \includegraphics[width=\textwidth]{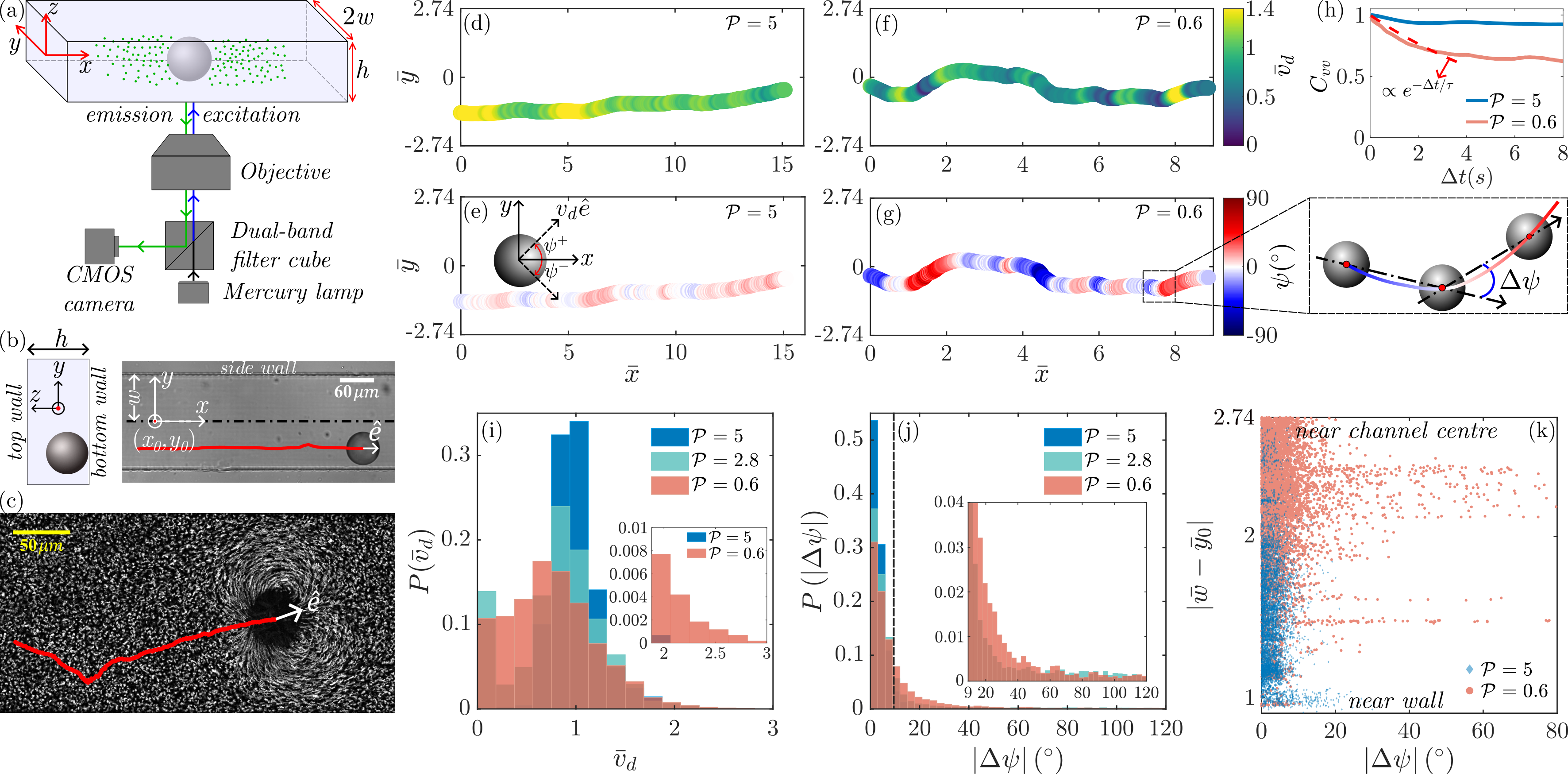}
\caption{\label{fig:1} \textbf{Run-and-tumble-like swimming of active droplets with increasing softness of microchannel walls}. \textbf{(a)} Schematic of the microscopy setup showing an active droplet in a microchannel with width $2w \approx 160$ $\mu$m and height $h \approx 100$ $\mu$m. 
\textbf{(b)} A bright field microscopy image of a self-propelling active droplet in a rigid microchannel. The trajectory of the droplet microswimmer is shown by the red line, and $\hat{e}$ represents the instantaneous swimming orientation. 
\textbf{(c)} A fluorescence micrograph depicting the trajectory and flow around an active droplet in a soft microchannel. Trajectory of an active droplet color-coded \textbf{(d)} with the non-dimensional instantaneous swimming speed $\bar{v}_d$, and \textbf{(e)} with the swimming orientation in the laboratory reference frame $\psi$ (as shown in inset) in a rigid microchannel $(\mathcal{P}=5)$.
\textbf{(f, g)} Trajectory of an active droplet swimming in a soft microchannel $(\mathcal{P}=0.6)$ color-coded with $\bar{v}_d$ and $\psi$ respectively. The blowup in \textbf{(g)} shows a schematic for a sharp re-orientation event during swimming in a soft microchannel.  
\textbf{(h)} Profiles of the velocity autocorrelation function $C_{vv} (\Delta t)$ for rigid and soft microchannels. 
\textbf{(i)} Distributions of $\bar{v}_d$, and \textbf{(j)} the instantaneous reorientation angle $\left| \Delta \psi \right|$ (blowup in (g)) for active droplets in microchannels with increasing softness (decreasing $\mathcal{P}$).
\textbf{(k)} Scatter plot showing the distribution of the distance of the droplet microswimmer centroid from the microchannel side walls (shown in (b)) for rigid and soft microchannels.}
\end{figure*}

In a microchannel with rigid walls $(\mathcal{P}=5)$, the active droplet swims at an approximately steady speed $(\bar{v}_d=\frac{|\vec{v}_d|}{v_0})$ along one of the side walls (Figs. ~\ref{fig:1}(b) and (d)).
The swimming orientation in the laboratory reference frame $\hat{e}$, as given by the angle $\psi$ (Fig. ~\ref{fig:1}(e) inset), also remains approximately parallel to the wall (Fig. ~\ref{fig:1}(e)).
This unidirectional ballistic swimming along the wall of a microchannel is intrinsic to these droplet microswimmers \cite{dey2022oscillatory, buness2024electrotaxis}.  
However, in a soft microchannel $(\mathcal{P}=0.6)$ of identical geometry, the active droplet swims with a smaller speed, and intermittently slows down, often stops and subsequently speeds up (Fig. ~\ref{fig:1} (f); SM- Video S1 and S2 \cite{sontakke2025_supplemental}).
Interestingly, unlike for $\mathcal{P}=5$, there are frequent and significant changes in $\psi$ for $\mathcal{P}=0.6$ (Figs. ~\ref{fig:1}(c) and (g); see SM-Fig. \ref{fgr: SM4} for more examples of similar trajectories \cite{sontakke2025_supplemental}), reminiscent of `tumbling' in biological microswimmers, like flagellated bacteria \cite{lauga2016bacterial}.
This loss of unidirectional swimming is evident from the variations of the velocity autocorrelation function $C_{vv}(\Delta t)=\langle \frac{\vec{v}_d(t+\Delta t) \cdot \vec{v}_d(t)}{|\vec{v}_d(t+\Delta t) ||\vec{v}_d(t)|}\rangle_t$ (Fig. ~\ref{fig:1}(h)).
While $C_{vv}$ for $\mathcal{P}=5$ remains constant about 1 with increasing time interval $\Delta t$, it decays with $\Delta t$ for $\mathcal{P}=0.6$ (Fig. ~\ref{fig:1}(h)).
An exponential fit to the initial decay, $C_{vv}(\Delta t) \propto e^{-\Delta t/ \tau}$ (red dashed line in Fig. ~\ref{fig:1}(h)) gives an analogous persistent (`run') length of $\bar{l}_p \sim 4.6$ for the microswimmer in the soft microchannel.
Here, $\tau \sim (l_p/\langle v_d \rangle_{\mathcal{P}=0.6})$, and $\langle v_d \rangle_{\mathcal{P}=0.6} = 18.04$ $\mu$m/s is the average swimming speed for $\mathcal{P}=0.6$.  
Such reorientations in the swimming direction result in a reduction $(\sim 10 \%)$ in the slope of the MSD plot at long time, $\propto \Delta t^{1.8}$, from the classical ballistic behaviour, $ \propto \Delta t^{2.0}$, that is observed for $\mathcal{P}=5$ (see SM-Fig. \ref{fgr: SM6} \cite{sontakke2025_supplemental}).

To further substantiate the key features of the emergent swimming dynamics in soft microchannels, we plot the probability distributions for $\bar{v}_d$ (Fig. ~\ref{fig:1}(i)), and for the magnitude of the reorientation angle $|\Delta \psi|$ (Fig. ~\ref{fig:1}(j)) using about 25 experimental runs consisting of 10214 data points. 
$\left|\Delta \psi \right|(t_i) =| \psi (t_i+\Delta t) - \psi(t_i)|$ is defined as the magnitude of the difference in $\psi$ between two consecutive time instants (blowup in Fig. ~\ref{fig:1}(g)).   
For $\mathcal{P}=5$, the $P(\bar{v}_d)$ distribution (Fig. ~\ref{fig:1}(i)) is narrow, and symmetric about the peak at the average swimming speed (by definition $\langle \bar{v}_d \rangle_{\mathcal{P}=5}=\frac{\langle {v}_d \rangle_{\mathcal{P}=5}}{v_0}=1$) representing steady swimming. 
With increasing softness of the microchannel, $P(\bar{v}_d)$ becomes broader and asymmetric about the peak, which shifts to lower values of $\bar{v}_d$.
For $\mathcal{P}=0.6$, on the one hand, the greater occurrence of smaller swimming speeds represents slower runs and intermittent slowing of the microswimmer, often including momentary stalling. 
On the other hand, the extended tail of the $P(\bar{v}_d)$ distribution over relatively larger speeds $(\bar{v}_d \sim 2-3)$ represents the intermittent acceleration (Fig. ~\ref{fig:1}(i) and its inset).
For $\mathcal{P}=5$, the reorientation events are weak with an average value of $\Delta \psi \approx 9$ $^\circ$ (black dashed line in Fig. ~\ref{fig:1}(j); also see SM-Fig. \ref{fgr:SM5} \cite{sontakke2025_supplemental}). 
With a reduction in $\mathcal{P}$, $P(\Delta \psi \approx 0 \; ^\circ)$ reduces, indicating the loss of unidirectional swimming. 
Furthermore, in the softer microchannels, events with $\Delta \psi > 9 \; ^\circ$ (Fig. ~\ref{fig:1}(j) and its inset), indicate the frequent reorientaion or the `tumbling' events.  
These sharp reorientation events in the softer microchannels alter the usual wall-hugging swimming of the droplet microswimmers, and these mostly swim away from the side walls (i.e. the walls of the microchannel perpendicular to the $X$-$Y$ plane; Fig. \ref{fig:1}(b)).
This can be inferred from the scatter plot (Fig. ~\ref{fig:1}(k)) for the transverse distance of the microswimmer centroid from one of the side walls ($\left| \bar{w} - \bar{y_{0}} \right|$) versus the corresponding $\left| \Delta \psi \right|$ for both rigid and soft microchannels. 
Note that the emergent run-and-tumble-like swimming in the soft microchannels is not  triggered by any change in the interfacial chemistry of the droplet microswimmer due to the PDMS prepolymer in the softer microchannels. 
This can be inferred from the fact that unidirectional swimming along the wall is recovered in the rigid part of a microchannel having spatially varying wall elasticity, as the active droplet swims from the soft side to the rigid side (see SM- Fig. \ref{fgr:SM7} and Video S3 \cite{sontakke2025_supplemental}).

\subsection*{Correlation between reorientation in swimming direction (`tumbling') and local variation in swimming speed} 
\begin{figure}[ht]
 \centering
 \includegraphics[width=0.48\textwidth]{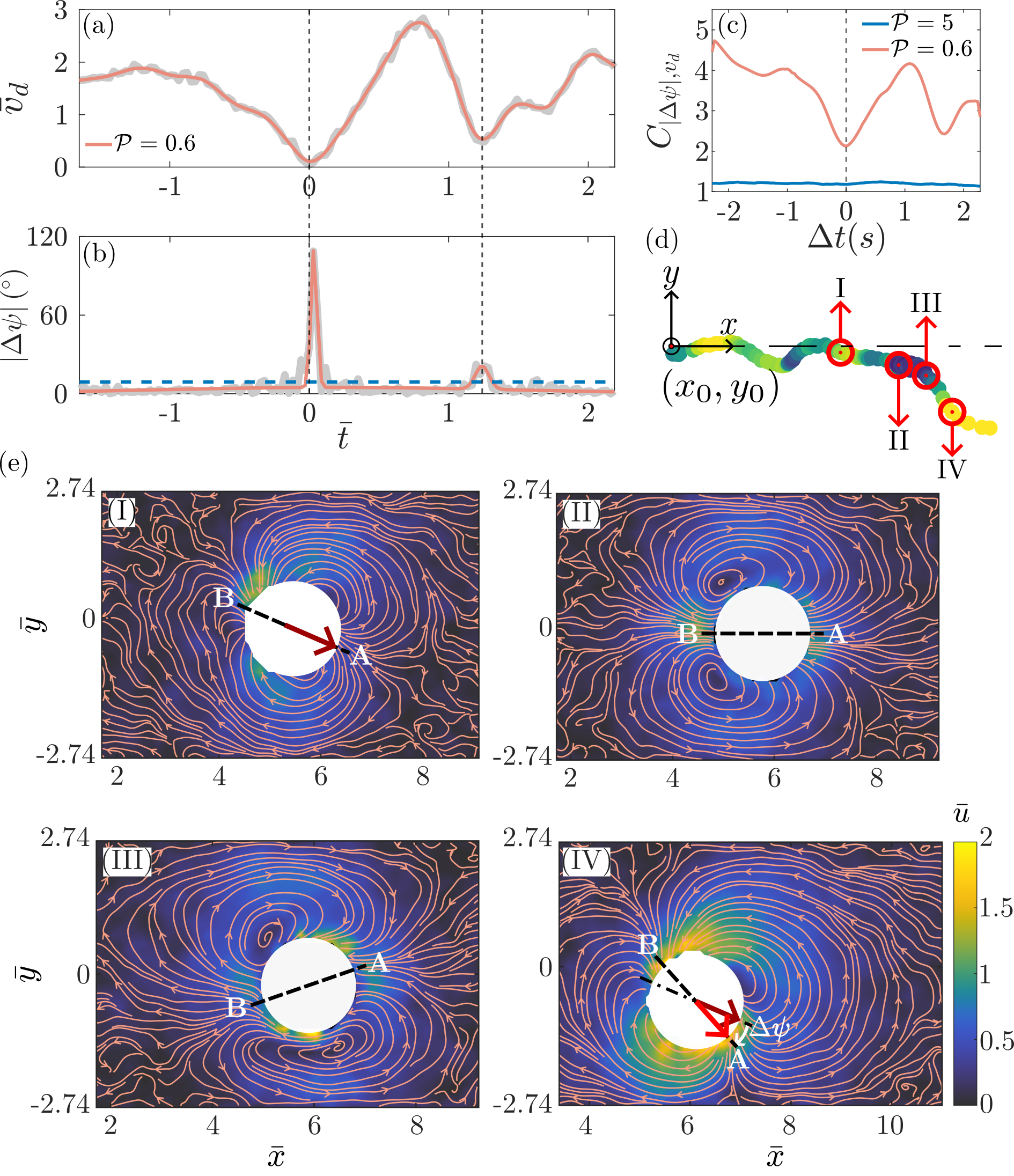}
\caption{\label{fig:2} \textbf{Correlation between sharp reorientation in swimming direction and local variation in swimming speed and velocity field in a soft microchannel.}\textbf{(a)} Temporal variation of the non-dimensional swimming speed $\bar{v}_d$ for an active droplet in a soft microchannel $(\mathcal{P}=0.6)$, and \textbf{(b)} the corresponding variation in the reorientation angle $\Delta \psi$ (faded solid line: experimental data; bold solid line: fit to the experimental data). The mean value for $\Delta \psi$ in a rigid microchannel $(\mathcal{P}=5)$ is marked with the blue dotted line. \textbf{(c)} Correlation function between the local reorientation angle $\Delta \psi$ and the swimming speed ${v}_d$ for rigid and soft microchannels.  \textbf{(d)} Swimming trajectory of an active droplet in a soft microchannel $(\mathcal{P}=0.6)$ color-coded with $\bar{v}_d$.  \textbf{(e)} Changes in the hydrodynamic signature (represented by streamlines and local flow velocity $(\bar{u})$ contour plots) exhibited by the droplet microswimmer at instances marked by I, II, III, and IV in (d), during a sharp reorientation event.}
\end{figure}

Interestingly, the sharp reorientations of the droplet microswimmer in the soft microchannels are preceded and followed by the local reduction and increase in swimming speed, respectively (typical examples are shown in Figs.~\ref{fig:2}(a) and (b)). 
Note that $\left| \Delta \psi \right|$ for the reorientation events shown for $\mathcal{P}=0.6$ are greater than the $\left| \Delta \psi \right|$ observed for $\mathcal{P}=5$ (blue dashed line in Fig. ~\ref{fig:2}(b) shows the average for $\mathcal{P}=5$; the exact variations are shown in SM-Fig. \ref{fgr:SM5} \cite{sontakke2025_supplemental}).
In Figs.~\ref{fig:2}(a) and (b), $\bar{t}=0$ marks the time instant at which $\bar{v}_d$ is minimum.
The peak in $\left| \Delta \psi \right|$, indicating the first sharp reorientation, immediately follows this instant, and subsequently the microswimmer speeds up over a unidirectional run until the next reorientation event (Figs. ~\ref{fig:2}(a) and (b)).
This characteristic sequence of decelaration, reorientation, and subsequent accelaration is repeated at successive reorientation events in the soft microchannels. 
Note that the active droplet does not always necessarily come to a perfect stop during a reorientation event, but its swimming speed considerably decreases, almost to zero, just prior to the reorientation.  
Since the local minima in $\bar{v}_d$ coincide with the beginning of the peaks in $\left| \Delta \psi \right|$, the two quantities are negatively correlated for the swimming dynamics in the soft microchannels.
This is indicated by the local dips in the cross-correlation function $C_{\left|\Delta \psi \right|, v_d}$ for $\mathcal{P}=0.6$ (Fig. ~\ref{fig:2}(c)).
There is no such correlation between $v_d$ and $\left| \Delta \psi \right|$ for $\mathcal{P}=5$ (Fig. ~\ref{fig:2}(c)).

\subsection*{Variation in hydrodynamic signature over a sharp reorientation event} 
The autonomous changes in the hydrodynamic signature of the droplet microswimmer during a typical sharp reorientation event, as discussed above, are highlighted in Figs. \ref{fig:2}(eI-eIV) (also see Video S2 \cite{sontakke2025_supplemental}).
The instants along the trajectory at which the velocity fields are shown in Figs. \ref{fig:2}(eI-eIV), are marked (red circles) in Fig. \ref{fig:2}(d).
For $\mathcal{P}=0.6$, the active droplet swims by generating a dipolar velocity field (Fig. \ref{fig:2}(eI)), as has also been witnessed in rigid micro-confinements \cite{guchhait2025flow}.
As the microswimmer comes to a momentary halt (no translation), the persistence of the interfacial activity still generates a velocity field around the droplet characterized by a pair of vortices at the posterior (Fig. \ref{fig:2}(eII)).
Note that when the droplet is almost stationary, anterior refers to the droplet hemisphere containing point A of the line A-B, and when the droplet is swimming, it refers to the hemisphere facing $+\hat{e}$. 
During this momentary halt, the active droplet squirms about its location as highlighted by the reorientation of the posterior vortex pair (Fig. \ref{fig:2}(eIII)).
Eventually, the droplet microswimmer reorients and starts to swim with a dipolar flow field in a direction (red arrow in Fig. \ref{fig:2}(eIV)) significantly different from its initial swimming orientation (maroon arrow in Figs. \ref{fig:2}(eIV) and (eI)), and thereby undergoes a sharp reorientation.
The instantaneous acceleration of the microswimmer can be inferred from the increase in the local flow velocity $(\bar{u})$ around the droplet (compare the contour plots of Figs. \ref{fig:2}(eIV) and (eI)), as well as in the swimming speed in Fig. \ref{fig:2}(d).

\section{Understanding the emergence of run-and-tumble-like swimming in soft microchannels}
\subsection*{Self-propelling microswimmers slow down in soft microchannels due to elastohydrodynamic interactions}
\begin{figure}[ht]
 \centering
 \includegraphics[width=0.48\textwidth]{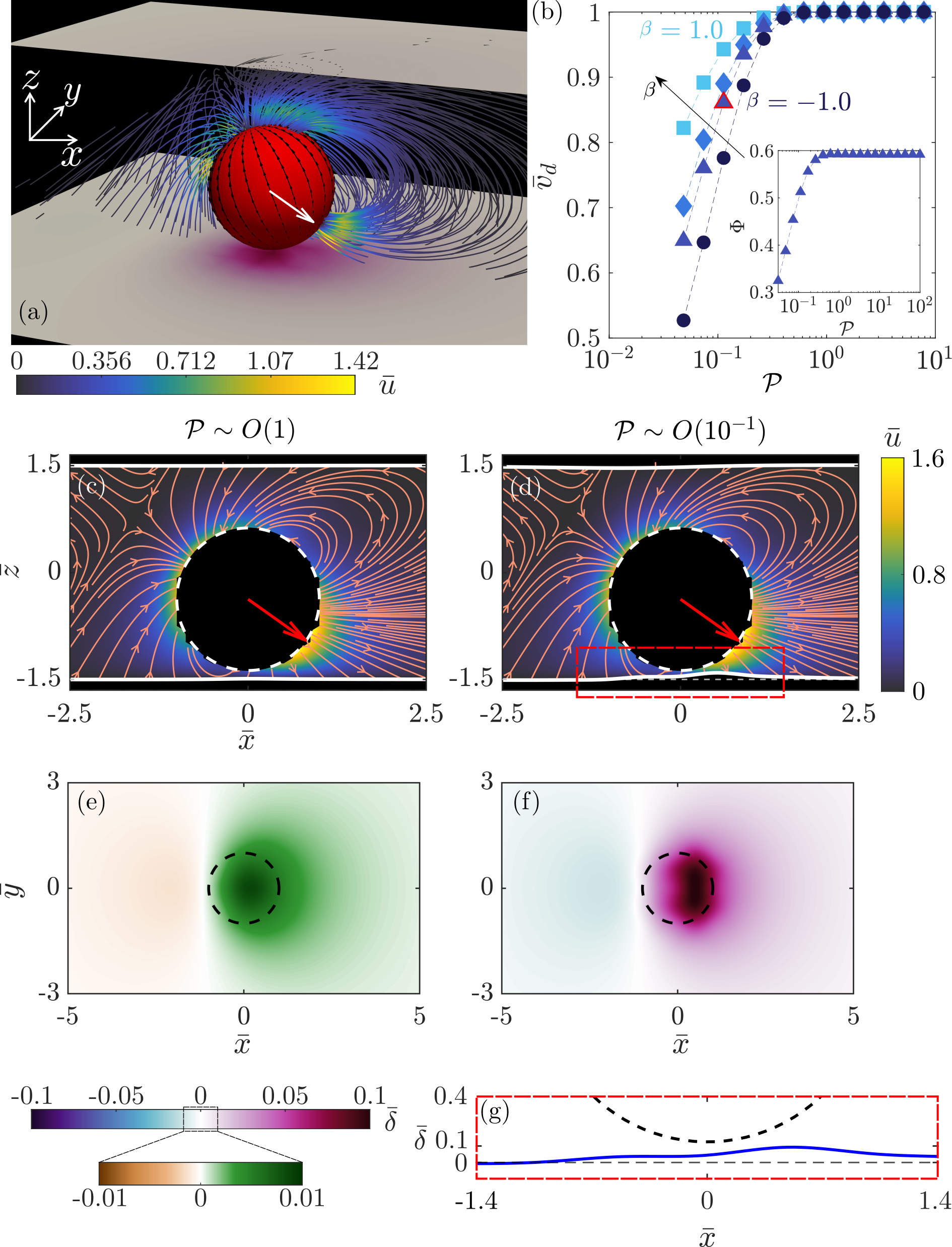}
\caption{\label{fig:3} \textbf{3D numerical simulation results showing the reduction in swimming speed of a self-propelling microswimmer due to wall deformation in a soft microchannel.} 
(\textbf{a}) 3D velocity field generated by a spherical, pusher-type squirmer ($\beta = -0.5$) in a soft microchannel ($\mathcal{P} \sim \mathcal{O}(0.1)$; non-dimensional wall separation distance is 3).
The deformation ($\bar{\delta}$) of the neighbouring soft wall induced by the squirmer is represented using colour contours. 
(\textbf{b}) Variation in steady-state swimming speed of the microswimmer ($\bar{v}_d$) with the stiffness ($\mathcal{P}$) of the confining walls for different values of the squirmer parameter ($\beta=-1.0, -0.5, 0.5, 1.0$).
The inset in (b) shows the decrease in the swimming efficiency $(\Phi)$ for the squirmer $(\beta=-0.5)$ with reducing $\mathcal{P}$.
Velocity field generated by the squirmer ($\beta = -0.5$) in the $X-Z$ plane while swimming in (\textbf{c}) a rigid microchannel ($\mathcal{P} \sim \mathcal{O}(1)$), and (\textbf{d}) in a soft microchannel ($\mathcal{P} \sim \mathcal{O}(0.1)$). The solid line represents the deformation profile for the adjacent wall, while the dashed line represents the undeformed wall position. 
Deformation contour plots of the bottom wall perpendicular (parallel) to the $X-Z$ ($X-Y$) plane for (\textbf{e}) $\mathcal{P} \sim \mathcal{O}(1)$ and (\textbf{f}) $\mathcal{P} \sim \mathcal{O}(0.1)$ (dotted circle shows the projection of the squirmer on the plane).(\textbf{g}) Blowup of the wall deformation ($\bar{\delta}$) profile (shown in \textbf{d}).}
\end{figure}

To explain the emergence of the aforementioned run-and-tumble-like swimming, we first study the effect of channel stiffness on steady-state swimming speed of the droplet microswimmer. 
For this, we perform 3D simulations of microswimmers in soft microchannels based on the boundary integral method (Fig. \ref{fig:3}(a)).
The self-propelling active droplet is modeled as a pusher-type, spherical squirmer \cite{lauga2020fluid}.
The spherical squirmer is characterized by the squirmer parameter $\beta$ ($<0$ for pushers), which determines the distribution of the slip velocity on its surface (Fig. \ref{fig:3}(a)).
Interestingly, the simulation results show that the steady-state value of $\bar{v}_d$ reduces with the reducing value of $\mathcal{P}$ (Fig. \ref{fig:3}(b)).
Note that the reduction in $\bar{v}_d$ in a microchannel of definite softness depends on the size of the confinement relative to $R_d$ (see SM- Fig. \ref{fgr:SM8} \cite{sontakke2025_supplemental}).
Specifically, the experimentally observed reduction in $\left< \bar{v}_d \right>$ to $\sim 0.8$ on reducing $\mathcal{P}$ from $\sim \mathcal{O}(1)$ to $\sim \mathcal{O}(0.1)$  (Fig. \ref{fig:1}(i))  is observed in the simulations for $\beta=-0.5$ and for non-dimensional wall separation distance of 3 (Fig. \ref{fig:3}(b)).
Hence, we conclude that for the droplet microswimmer in the microchannel (Fig. \ref{fig:1}(b)), the confinement along the $Z-$direction ($X-Z$ plane; $\bar{h} \approx 3$) plays a dominant role in slowing the microswimmer compared to the confinement along the $Y-$direction ($X-Y$ plane; $2\bar{w} \approx 6$; SM- Fig. \ref{fgr:SM8}(c) \cite{sontakke2025_supplemental}).

For $\mathcal{P} \sim \mathcal{O}(0.1)$, surface traction vector corresponding to the velocity field generated by the droplet microswimmer (Fig. \ref{fig:3}(d)) results in the upward deformation $(\bar{\delta}=\delta/R_d; \bar{\delta}>0)$ of the soft wall of the microchannel immediately underneath it, i.e., the bottom wall perpendicular (parallel) to the $X-Z$ $(X-Y)$ plane (Figs. \ref{fig:3}(f) and (g)).
The deformation of the soft wall towards the microswimmer is larger underneath it, specifically on the anterior side, and gradually decays away from it (Figs. \ref{fig:3}(f) and (g)). 
There is a smaller downward (negative) displacement of the soft wall at the posterior of the microswimmer (Fig. \ref{fig:3}(f)).
Note that wall deformation is relatively negligible for an identical microswimmer in the rigid microchannel $\mathcal{P} \sim \mathcal{O}(1)$ (Figs. \ref{fig:3}(c) and (e)).
The deformation of the soft wall underneath the microswimmer significantly reduces the thickness of the intervening thin liquid film (Fig. \ref{fig:3}(g)).
This results in increased lubrication pressure in the thinner liquid film.
Hence, for a fixed surface velocity distribution (fixed $\beta$), it becomes difficult for the squirmer to swim with the same speed by transporting liquid from the anterior to the posterior through the thinner film.
Consequently, $\left< \bar{v}_d \right>$ reduces for the microswimmer in the soft microchannel with deformable walls.
The swimming efficiency (Lighthill efficiency, $\Phi$) of the microswimmer also reduces with reducing $\mathcal{P}$ (inset in Fig. \ref{fig:3}(b)).  

\subsection*{Growth of the chemical coverage around a slowly moving active droplet initiates `tumbling' in soft microchannels}
\begin{figure}[h!]
 \centering
 \includegraphics[width=0.48\textwidth]{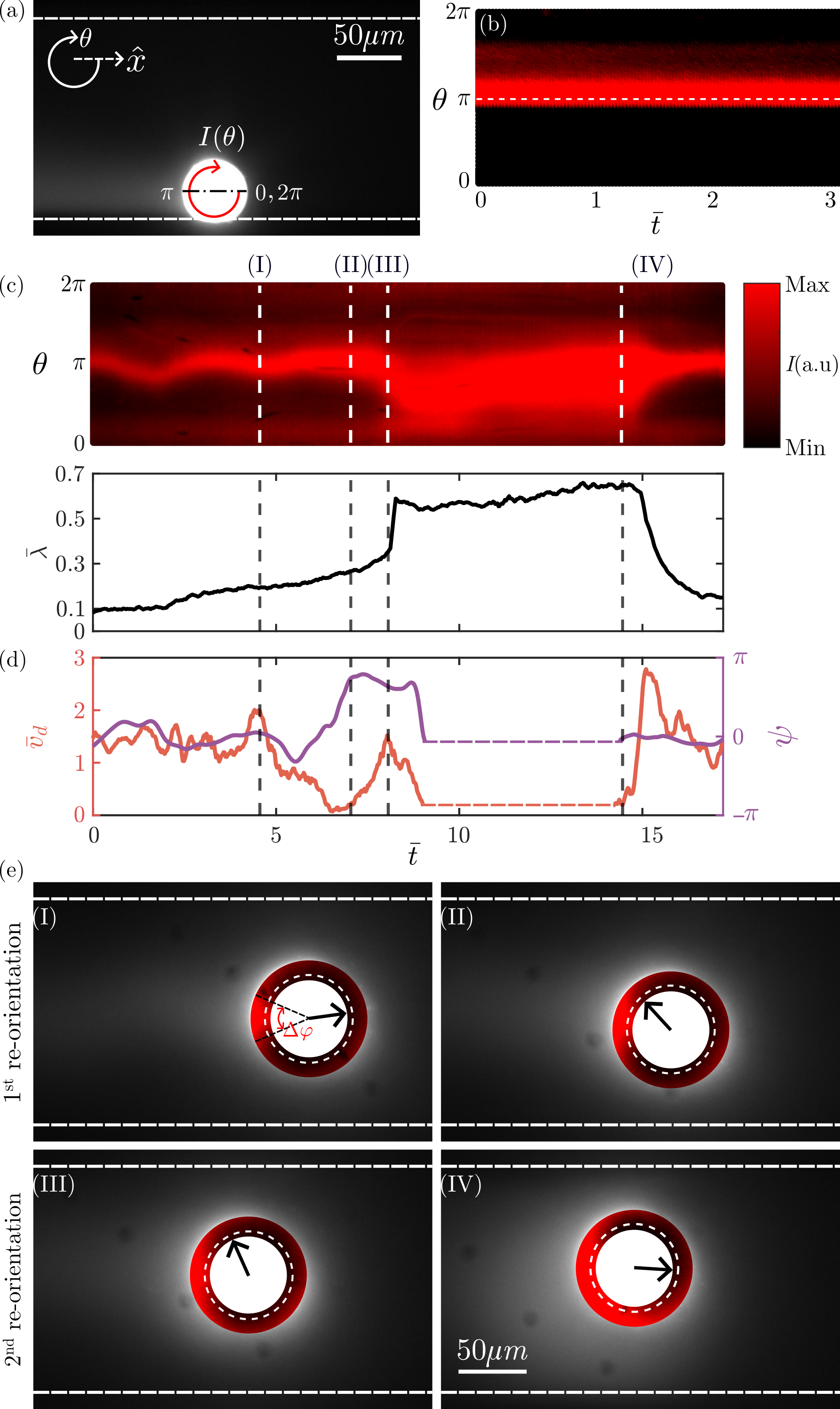}
\caption{\label{fig:4}\textbf{Evolution of the filled micelle concentration around the active droplet over local reorientation events}. \textbf{(a)} A fluorescence micrograph showing the filled micelle trail behind an active droplet in a rigid microchannel $(\mathcal{P}=5)$. 
\textbf{(b)} Corresponding kymograph of fluorescence intensity $I(\theta)$ (filled micelle concentration) around the active droplet during its self-propulsion. 
\textbf{(c)} Kymograph of $I$ for the active droplet in the soft microchannel $(\mathcal{P}=0.6)$ over two re-orientation events; temporal variation of the fractional coverage of the microswimmer's periphery by the filled micelle wake ($\bar{\lambda}(t)=\Delta \varphi(t)/2 \pi$) is also shown here.
\textbf{(d)} The corresponding variations in swimming speed $(\bar{v}_d)$ and swimming orientation $(\psi)$, over two reorientation events demarcated by instants I-II and III-IV respectively.
\textbf{(e, I-IV)} Fluorescence micrograph sequence showing the variation in the filled micelle concentration around the active droplet, and its instantaneous swimming orientation, over the reorientation events. The time instances shown in (e, I-IV) are the ones marked in the kymograph and in the temporal variations of $\bar{v}_d$ and $(\psi)$ by dotted lines.}
\end{figure}

As the active droplet swims, it leaves behind a trail of self-generated chemical (filled micelle) whose temporal evolution can be visualized by constructing a kymograph \cite{hokmabad2021emergence, hokmabad2022chemotactic}.
For $\mathcal{P}=5$, the chemical accumulates about the posterior apex of the microswimmer which is diametrically opposite to the swimming orientation (in this case, $\hat{e}$ and $+\hat{x}$ are identical, and hence, the posterior apex is about $\theta \sim \pi$ relative to $+\hat{x}$) (Figs. \ref{fig:4}(a) and (b)).
Furthermore, the chemical trail remains uniform in thickness over the swimming trajectory (Fig. \ref{fig:4}(b)).
Fig. \ref{fig:4}(c) shows the corresponding kymograph over two representative reorientation events (over instants I-II and III-IV respectively) for the active droplet swimming in the soft microchannel $(\mathcal{P}=0.6)$.
This includes one reorientation event during which the active droplet momentarily stalls (2$^{\text{nd}}$ reorientation; demarcated by instants III-IV).
Points II and IV in Fig. \ref{fig:4}(c) represent the instants at the beginning of a new `run' after a `tumble'.  
As the droplet microswimmer slows down due to the aforementioned elastohydrodynamic interactions (Figs. \ref{fig:3}(d) and (g)), the local velocity field (Fig. \ref{fig:2}(e)) due to the interfacial activity continues to advect the filled micelles around it.
The advection-dominated transport of the chemical around the periphery of the slowly moving droplet microswimmer results in gradual spreading of the chemical wake about its posterior apex.
This further slows down the microswimmer. 
Note the growth of the chemical wake in the kymograph, as quantified by $\bar{\lambda}(t)=\Delta \varphi(t)/2 \pi$ (Figs. \ref{fig:4}(c) and (eI)), vis-a-vis the reduction in $\bar{v}_d$ (Fig. \ref{fig:4}(d)) between instants I-II and III-IV.
Even during a momentary stalling, the local vortex pair (Figs. \ref{fig:2}(e)II, III) continues to advect the filled micelles around the microswimmer, resulting in the growth of the chemical coverage (Fig. \ref{fig:4}(c) between instants III-IV).
As the filled micelle coverage grows around the almost stationary droplet microswimmer, it eventually gets pulled by the gradient of empty micelles (surfactants) along a different direction uncovered by filled micelles.
This results in a re-orientation in the $X-Y$ plane and a ballistic motion (`run') along the new direction (note the variations in $\bar{v}_d$ and $\psi$ following the reorientation, i.e. instants II and IV onwards).
The growth of the chemical wake along the periphery of the droplet microswimmer over a reorientation, and the corresponding change in the swimming orientation, can also be inferred from the fluorescence micrographs in Fig. \ref{fig:4}(e). 
Here, I and II, for the $1^{st}$ reorientation, and III and IV, for the $2^{nd}$ reorientation, represent the identical instants marked in Figs. \ref{fig:4}(c) and (d).    
The colour band around the droplet microswimmer represents the instantaneous concentration of the filled micelle along its circumference, and the black arrow marks the instantaneous orientation of $\hat{e}$.
The aforementioned coupling between the elastohydrodynamic interaction induced slowing down of the microswimmer, and its interaction with the self-generated chemical wake explains not only the emergence of the reorientations, but also why a reorientation event is preceded and followed by local decrease and increase in $\bar{v}_d$ respectively (Figs. \ref{fig:2}(a) and (b)).
The important role of the advection-driven transport of the chemical in the reorientation dynamics also explains why the microswimmer reorientation typically takes place over the advective time scale i.e. $\bar{t} \sim \mathcal{O}(1)$ (Figs. \ref{fig:2}(a) and (b)).  

\section{Conclusion}
In this work, we demonstrate and explain the emergence of run-and-tumble-like motility for autophoretic microswimmers, like active droplets, in soft microchannels.
While in a rigid microchannel the active droplet swims following a unidirectional trajectory with a steady speed along a wall, in a soft microchannel it undergoes frequent reorientations, reminiscent of `tumbling' in biological microswimmers, and thereby enhancing its ability for spatial exploration. 
Such reorientation events are preceded and followed by local decrease and increase in swimming speed, respectively.
We explain these alterations in swimming characteristics in response to the increase in the softness of the microchannel walls using microscopy experiments and 3D boundary integral simulations.
We numerically show that the near-wall elastohydrodynamic interactions, induced by the microswimmer flow field in the narrow confinement, result in deformation of the soft wall adjacent to the microswimmer.
For a fixed swimming strength, the deformation of the soft wall hinders the fore-to-aft transport of liquid through the thinner lubrication film. 
This culminates in the slowing down of the self-propelling microswimmer, and in the reduction of its swimming efficiency.
Subsequently, the local growth of the self-generated chemical (filled micelles) wake around the slowly moving microswimmer results in further reduction in the swimming speed.
Eventually, the interaction of the slowed-down microswimmer with the gradient of empty micelles leads to a local reorientation in the swimming direction, and a subsequent `run'.

Although elastohydrodynamic interactions of passive particles near soft, deformable walls are well established \cite{skotheim2004soft, saintyves2016self, essink2021regimes, bertin2022soft}, the same has remained poorly understood for active particles, like self-propelling microswimmers.
Furthermore, the fact that autophoretic microswimmers respond to increasing softness of the narrow confinement walls by autonomously altering their motility has never been demonstrated before.
We show that the coupling between the elastohydrodynamic interactions and the chemo-hydrodynamics, inherent in the self-propulsion mechanism, remarkably results in an emergent run-and-tumble-like swimming dynamics for autophoretic microswimmers in soft microchannels.
This altered motility state, stemming from a combination of elastic and chemo-hydrodynamic interactions, has hitherto remained unknown and can have far-reaching implications.

The spontaneous adaptation of motility by the autophoretic microswimmer to changes in the elasticity of microchannel walls (a fact that can also be inferred from Video S3 \cite{sontakke2025_supplemental}) imparts to these microswimmers the ability to `sense' elasticity gradients, which has never been proposed before for artificial active systems.
We envisage that such `sensing' of variations in elasticity can result in a new `decision-making' capability of the synthetic microswimmers, which in turn can be exploited for tuning collective behaviour and for applications involving active sorting of microswimmers. 
These constitute interesting future studies, which our group is currently working on.
Finally, the elastohydrodynamic interactions of self-propelling microswimmers in soft microconfinements, as explained here, are valid for general squirmer-type microswimmers.
Hence, these results should have implications even for biological microswimmers. 
In this context, new experiments should be designed to probe the effects of such elastohydrodynamic interactions on the near-wall dynamics of biological microswimmers, specifically bacteria/algae, leading to surface attachment.  

\section*{Acknowledgements}
S.S.S. acknowledges the Prime Minister’s Research Fellowship (PMRF), a scheme by the Govt. of India to improve the quality of research in various research institutions in the country. R.D. acknowledges support from Science and Engineering Research Board (SERB) (now subsumed under Anusandhan National Research Foundation), Department of Science and Technology (DST), Government of India, through Grant No. SRG/2021/000892. R.D. also acknowledges support from IIT Hyderabad through seed Grant No. SG 93. M.S.R acknowledges support from Science and Engineering Research Board (SERB) (now subsumed under Anusandhan National Research Foundation), Department of Science and Technology (DST), Government of India, through Grant No. SRG/2021/001020.  

\bibliography{Bibliography}
\newpage
\clearpage
\widetext 
\renewcommand{\thefigure}{S\arabic{figure}}
\begin{center}
    
    {\large \bfseries Supplemental material for Emergence of run-and-tumble-like swimming in self-propelling artificial swimmers in soft microchannels}\\[2ex]

 \textit{Smita S. Sontakke}$^{1}$,
    \textit{Aneesha Kajampady}$^{1}$,
    \textit{Mohd Suhail Rizvi}$^{2}$,
    \textit{Ranabir Dey}$^{1}$ \\
    
    \textit{$^{1}$Department of Mechanical and Aerospace Engineering,}\\
    \textit{Indian Institute of Technology Hyderabad, Kandi, Sangareddy 502285, India} \\[1ex]
    \textit{$^{2}$Department of Biomedical Engineering,}\\
    \textit{Indian Institute of Technology Hyderabad, Kandi, Sangareddy 502285, India}
    
\end{center}

\section*{I. Experimental Materials and Methods}
\subsection*{1.1 Generation of active droplets}
\begin{figure*}[ht]
\centering
 \includegraphics[width=0.7\textwidth]{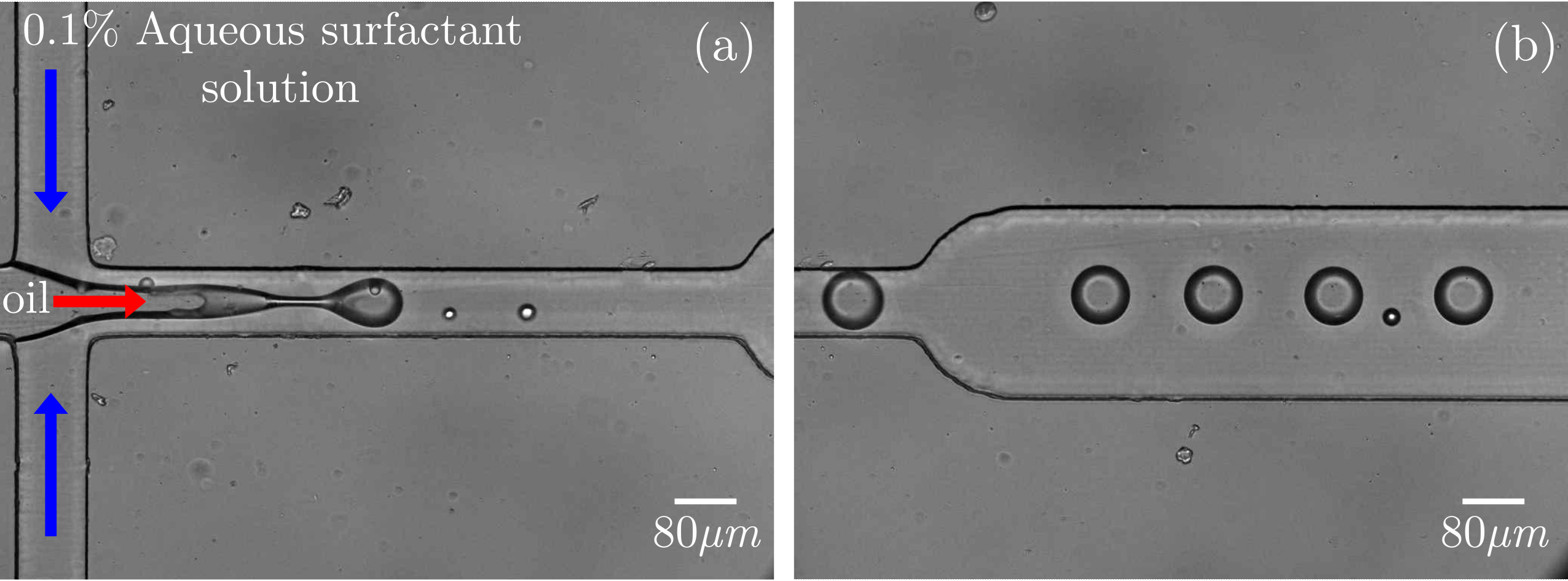}
 \caption {\label{fgr:SM1}Droplet generation using a flow-focusing device fabricated using photolithography. \textbf{(a)} The red and blue arrow indicates CB15 oil, 0.1 wt$\%$ TTAB solution, respectively, \textbf{(b)} Collection of monodisperse droplets in the reservoir at the outlet.}
\end{figure*}

We use a standard microfluidic flow focusing device fabricated using photolithography to generate the oil droplets (SM-Figs. \ref{fgr:SM1}(a) and (b)). We pour the PDMS solution prepared by mixing base polymer-to-cross-linker in a $10:1$ ratio on the photo-lithographic mould and cure it for 3 hours at $75^\circ$C in a hot air oven. We then puncture the two inlets and the outlet, and then we expose this device and a pre-cleaned cover slip to the air plasma using the Harric Plasma cleaner (PDC-32G). We then immediately bond the PDMS device to the cover slip. The two inlets for aqueous surfactant solution and the oil are then connected to the precision syringe pumps (F101 and F100X; Chemyx), and the outlet is connected to a glass vial to collect the droplets using the PTFE tubing. The two jets of $0.1\%$ TTAB solution in DI water with a flow rate of $\sim 100 \mu l/hr$ break the oil jet, having a flow rate of $\sim 0.2 \mu l/hr$ governed by the Rayleigh–Plateau instability (SM-Fig. \ref{fgr:SM1}(a)).

\subsection*{1.2 Fabrication of soft microchannel}
\begin{figure*}[ht]
\centering
 \includegraphics[width=0.7\textwidth]{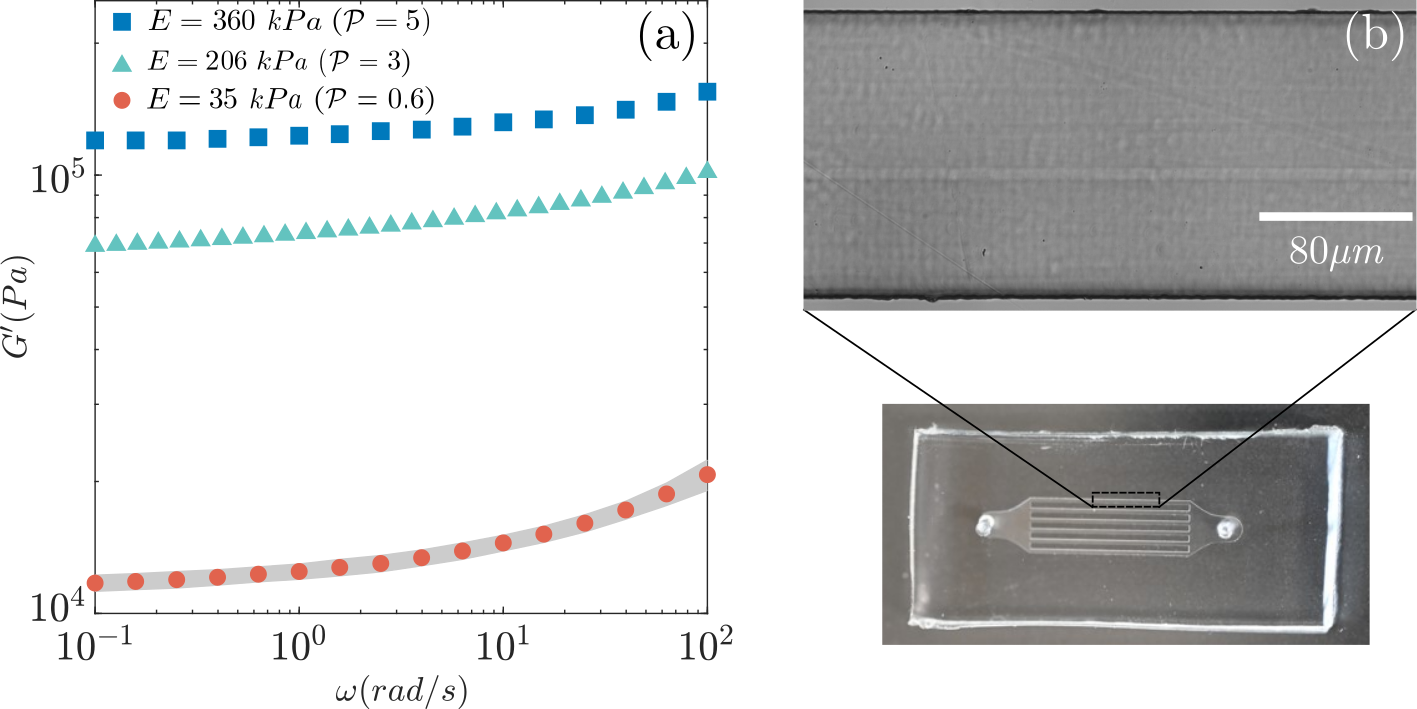}
 \caption{\label{fgr:SM2}\textbf{(a)} Storage modulus ($G^{\prime}$) for varying PDMS base polymer to cross-linker ratio, \textbf{(b)} Close-up view of the microchannels on a microfluidic chip bonded with a PDMS-coated coverslip.}
 \end{figure*}

The soft microchannels are fabricated by increasing the concentration of polydimethylsiloxane (PDMS) base-polymer in base polymer-to-crosslinker mixture (10:1, 25:1, and 40:1). To characterize the rheological properties such as storage modulus ($G^{\prime}$) and loss modulus ($G^{\prime\prime}$) of these PDMS samples with varying polymer-to-crosslinker ratios, we prepare the circular discs ($1''$ diameter and $\sim 1.5 mm$ thick) of these PDMS solutions and cure the sample with $10:1$ mixing ratio for 3 hours at $75^\circ$C and the samples with $25:1$ and $40:1$ mixing ratio require extended curing of 17 hours in a hot air oven at the same temperature. After 17 hours, the $25:1$ and $40:1$ PDMS cures as it peels off cleanly from the mould. To confirm complete curing at $40:1$ mixing ratio, additional tests are conducted after 30 and 40 hours of curing, revealing no significant changes in rheological properties beyond 17 hours. The storage modulus $G^{\prime}$, represents the elastic response of the cured polymer (SM-Fig. \ref{fgr:SM2}(a)).
The gray shaded area represents the standard deviation of the $40:1$ sample cured for 30 and 40 hours.
The storage modulus $G^{\prime}$ increases for higher cross-linker concentration, illustrating a stiffer/rigid and more elastic behavior. The Elastic response of the cured PDMS is measured in terms of Young's modulus $E$, calculated as $E = 2G^{\prime}(1+\nu)$; considering PDMS as a nearly incompressible material, the Poission's ratio $\nu$ is taken as 0.5.
We coat the cover slips (160 $\mu$ m thick) with the PDMS mixed in the same ratio as for the microchannel using a spin coater and cure it for 17 hours at $75^\circ$C.  We then expose the PDMS-coated glass cover slip and the cured PDMS stamp containing the negative replica of the mould to air plasma for 2 minutes and bond them immediately (Fig. \ref{fgr:SM2}(b)).

\subsection*{1.3 Quantification of the error in $\mu$-PIV analysis}
The errors in $\mu$-PIV can be minimized by selecting tracer particles that are neutrally buoyant and follow the flow streamlines with minimal velocity lag.
The response time of the tracer particles to the flow changes is calculated from Newton's second law as $t_{0} = \rho_{p} d_{p}^{2}/18 \mu_{f}$, where $\rho_{p}$ is the density and $d_{p}$ is the diameter of the tracer particles, $\mu_{f}$ is the dynamic viscosity of the fluid medium \cite{raffel2018particle}. 
 For the fluorescent particles of size $500$ $nm$ ($480\pm16$ $nm$, Thermo Fisher Scientific) with the density of $1055$ $kg/m^{3}$ and dynamic viscosity of the fluid of $1$ mPas, the velocity la is calculated to be $7.9\times10^{-3}\mu m/s$. 
 Stokes number, which is defined as the ratio of particle response time $t_o$ and advective time scale $l/u_o$ is calculated to be in the order of $10^{-7}$. The Stokes number must be below 0.1 to keep tracer-induced errors under $1\%$ \cite{tropea2007springer}.

Errors may also result from inherent uncertainties in the calculation of displacement in the digital PIV analysis.  
These errors include bias error ($\epsilon_{bias}$), representing measurement accuracy, and random error ($\epsilon_{rms}$), reflecting precision \cite{gui2002correlation}. These displacement uncertainties can propagate to the velocity measurements, which is given as $U_{v}=[(\frac{\Delta X}{\Delta t}U_{s})^{2}+(\frac{s}{\Delta t}U_{\Delta X})^{2}+(\frac{s \Delta X}{\Delta t^{2}}U_{\Delta t})^{2}]^{1/2}$ \cite{chandrala2016unsteady}, where $U_s$, $U_{\Delta X}$, and $U_{\Delta t}$ are uncertainties in calibration factor, displacement, and time, respectively.
In this study, $\Delta t = 0.04$ s with $U_{\Delta t} = 2 \times 10^{-6}$ s. The calibration factor $s$ for a $20\times$ objective is $0.280226$ $\mu m$/pixel, and the uncertainty in the measurement of the calibration factor $U_{s}$ is $0.002865$ $\mu m$/pixel. In the present work, particle displacement between 2 frames, $\Delta x$, for the maximum flow velocity in the experiments of 40 $\mu m/s$ is calculated to be $\sim 6$ pixels. The uncertainty in displacement measurement $U_{\Delta X} =(\epsilon_{bias}^2+\epsilon_{rms}^2)^{1/2}$ for a particle image of $3\times3$ pixels (with a $20\times$ objective to resolve $500$ $nm$  particle diameter) using a DFT multipass algorithm is taken as $0.02$ pixel from \cite{phdthesis}. The overall maximum error due to all these uncertainties, the uncertainty in the velocity measurement is found to be $\pm 0.4520$ $\mu m/s$, which is $\pm 1.13$ $\%$ of the maximum velocity. 
A detailed comparison of the magnitude of the flow field generated by the droplet and the magnitude of the Brownian motion is done to verify that the contribution of Brownian velocity is negligible in the flow field generated around the droplet microswimmer \cite{guchhait2025flow}.

 \subsection*{1.4 Calculation of kymograph}
\begin{figure*}[ht]
\centering
\includegraphics[width=0.6\textwidth]{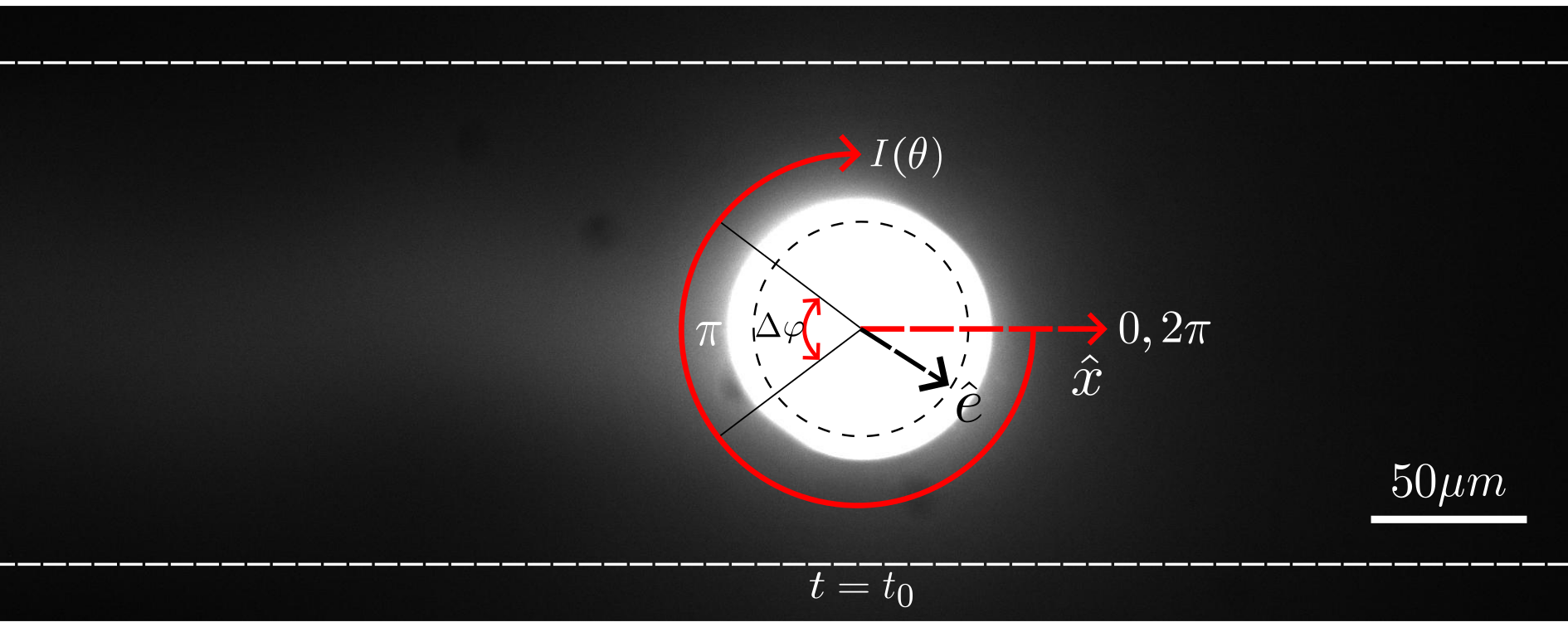}
 \caption{\label{fgr: SM3}Calculation of kymograph by extracting the fluorescence intensity around the droplet circumference $(0 \ to \ 2\pi)$ at a radial distance of $\sim$ $22 \mu m$ away from the circumference of the droplet.}
\end{figure*}

To analyze the fluorescence signal intensity around the droplet interface, we mix Nile Red dye (Invitrogen™) with the CB-15 oil and generate the fluorescence-tagged active droplets. Using an in-house developed MATLAB code, we extract the fluorescence intensity (Gray scale image 0-255) around the droplet ($0 - 2\pi$) at a distance of $22 \mu m$ away from the circumference of the droplet (SM-Fig. \ref{fgr: SM3}). We extract the fluorescence intensity $I(\theta)$ for an image sequence as the droplet is undergoing a run-and-tumble-like swimming trajectory. The $\theta$ is always increasing in a CW manner starting from the $+\hat{x}$. We then plot the kymograph that is the normalized intensities ($I/I_{max}$) over $\theta$ ranging from $0 - 2\pi$ on the ordinate and the time on the abscissa (see Figs. \ref{fig:4} (b) and (d) in the main text).  

 \subsection*{Calculation of thickness of chemical trail}
The thickness of the chemical trail generated around the droplet ($0 - 2\pi$) at a distance of $22 \mu m$ away from the circumference is quantified as the arc length $\bar{\lambda}(t) = (r_d+22) \Delta\varphi(t) / 2\pi (r_d+22)$ formed by the angle $\Delta\varphi$, where $r_d$ is the distance at which the intensity values are extracted. The angle $\theta$ is calculated based on the maximum intensity region at the posterior of the droplet.  
The intensities (Gray scale image 0-255) are non-dimensionalized using the maximum intensity $I_{max}$ around the droplet circumference across the entire image sequence. 
The region corresponding to maximum intensity is identified by applying a threshold based on $I_{max}$: only pixels with intensity values $\geq 50\%$ of $I_{max}$ are considered, while those with intensity values $< 50\%$ of $I_{max}$ are excluded. The angle $\Delta\varphi$ is then calculated based on the angular extent of the region satisfying this threshold. The non-dimensionalized arc length $\bar{\lambda}$ represents the percentage of maximum intensity area the droplet generates compared to its circumference.

\section*{II. Mathematical model and simulation}
\subsection*{2.1 Basic assumptions}
The modeling approach assumes that the squirmer swims in a viscous, incompressible fluid governed by the Stokes equations, appropriate for the low Reynolds number regime. 
The swimmer is confined within a rectangular channel where one lateral dimension (height) can be comparable to the squirmer diameter, while the other (width) is effectively unbounded, reducing the system to motion between two parallel walls. 
The squirmer propels itself via a prescribed tangential slip velocity on its surface, modeled using the classical axisymmetric squirmer formulation. 
The confining walls are treated as deformable elastic boundaries, and the domain beyond each wall is represented as a linearly elastic half-space, capturing the influence of substrate compliance on the fluid–structure interaction. 
No-slip conditions are imposed at the fluid–elastic interfaces, and the wall deformation is assumed to remain within the linear elastic regime. 
This framework allows the study of how confinement and substrate elasticity affect the squirmer's motion, flow field, and hydrodynamic interactions with boundaries.

\subsection*{2.2 Model}
\subsubsection*{2.2.1 Squirmer}
The swimmer, modeled as a spherical squirmer, generates propulsion through a prescribed tangential surface slip velocity. 
The surrounding fluid is Newtonian and governed by the Stokes equations in the limit of zero Reynolds number.
The fluid velocity $\mathbf{u}(\mathbf{x})$ and pressure $p(\mathbf{x})$ satisfy the steady incompressible Stokes equations
\begin{equation}
-\nabla p + \mu \nabla^2 \mathbf{u} = 0, \qquad \nabla \cdot \mathbf{u} = 0, \label{eq:stokes}
\end{equation}
where $\mu$ is the dynamic viscosity of the fluid.

Since the squirmer is confined between two parallel walls (Fig. \ref{fgr: SMnum}), and its surface slip velocity is axisymmetric about its propulsion axis, the system possesses geometric and rotational symmetry that simplifies the description of its configuration. 
Without loss of generality, we can specify the squirmer’s position in terms of its perpendicular distance in $z$ direction from one of the walls, effectively reducing the spatial degrees of freedom. 
Furthermore, the orientation of the squirmer relative to the wall can be fully characterized by a single angle, $\phi$, which denotes the angle between the swimmer’s propulsion axis and $x$-axis. 
This minimal parametrization, wall distance and orientation angle,captures the essential hydrodynamic interactions in the system while leveraging the axisymmetry of the swimmer and the translational invariance along the direction parallel to the walls.
On the surface of the squirmer, denoted $\partial D$, we impose the boundary condition
\begin{equation}
\mathbf{u}(\mathbf{x}) = \mathbf{U} + \boldsymbol{\Omega} \times (\mathbf{x} - \mathbf{x}_0) + \mathbf{u}_s(\mathbf{x}), \quad \text{for } \mathbf{x} \in \partial D,
\end{equation}
where $\mathbf{U}$ is the unknown translational velocity of the squirmer, $\boldsymbol{\Omega}$ is the unknown angular velocity, $\mathbf{x}_0$ is the origin of the swimmer's body frame (e.g., its center), and $\mathbf{u}_s(\mathbf{x})$ is the prescribed slip velocity defined over the squirmer’s surface.

The slip velocity $\mathbf{u}_s$ encodes the propulsive mechanism of the squirmer. 
For an axisymmetric squirmer, this is typically prescribed using the classical model:
\begin{equation}
\mathbf{u}_s(\theta) = B_1 \sin\theta\, \hat{\boldsymbol{\theta}} + B_2 \sin\theta \cos\theta\, \hat{\boldsymbol{\theta}},
\end{equation}
where $\theta$ is the polar angle relative to the swimmer's swimming axis, $\hat{\boldsymbol{\theta}}$ is the unit vector in the azimuthal direction on the surface, and $B_1$, $B_2$ are scalar coefficients. 
Here, $B_1$ determines the swimming speed for a squirmer in an unbound fluid, while the ratio $\beta = B_2/B_1$ characterizes the stresslet strength and hence the nature of the squirmer. 
If $\beta > 0$, the squirmer is a \emph{puller} (drawing fluid in along the swimming axis and ejecting it equatorially, e.g., algae like \textit{Chlamydomonas}); if $\beta < 0$, it is a \emph{pusher} (e.g., a bacterium like \textit{E. coli}); and if $\beta = 0$, the squirmer is \emph{neutral}.

The solution to equation \ref{eq:stokes} in an unbounded domain can be expressed using a boundary integral formulation. 
Specifically, the velocity at any point $\mathbf{x}$ in the fluid can be written in terms of surface tractions $\mathbf{f}$ acting on the boundaries $\partial D \cup B$ as:
\begin{equation}
u_i(\mathbf{x}) = -\frac{1}{8\pi\mu} \int_{\partial D \cup B} S_{ij}(\mathbf{x}, \mathbf{X}) f_j(\mathbf{X})\, dS(\mathbf{X}),
\label{eq:bie}
\end{equation}
where $S_{ij}(\mathbf{x}, \mathbf{X})$ is the free-space Stokeslet (Green's function) given by
\begin{equation}
S_{ij}(\mathbf{x}, \mathbf{X}) = \frac{\delta_{ij}}{r} + \frac{r_i r_j}{r^3}, \quad \text{with } \mathbf{r} = \mathbf{x} - \mathbf{X}, \; r = \|\mathbf{r}\|.
\end{equation}
This kernel is singular as $\mathbf{x} \rightarrow \mathbf{X}$, which complicates numerical evaluation of the integral, especially when $\mathbf{x}$ lies on or near the surface of integration.
Therefore, we need extra care in evaluating the boundary integral (see below).

\subsubsection*{2.2.2 Deformable boundry}

To incorporate substrate elasticity into the swimmer–boundary interaction, we model the confining walls of the channel as linear elastic half-spaces. In the classical Boussinesq problem, the vertical displacement field at the surface of an elastic half-space due to a point force applied normal to the surface is analytically tractable and forms the basis for our formulation. Specifically, the half-space is assumed to be isotropic, homogeneous, and linearly elastic, characterized by the Young's modulus $E$ and Poisson's ratio $\nu$. The surface deformation arises in response to the hydrodynamic tractions exerted by the fluid on the wall.

\begin{figure*}[ht]
\centering
\includegraphics[width=0.4\textwidth]{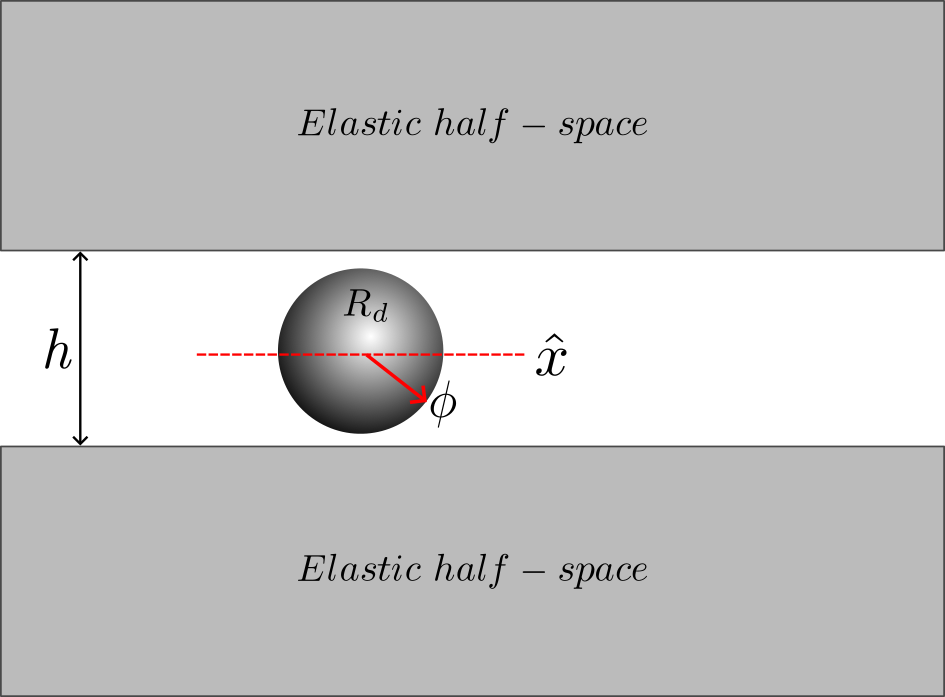}
 \caption{\label{fgr: SMnum}Schematic for numerical simulation.}
\end{figure*}

We consider the swimmer confined between two parallel, deformable walls separated by a distance $h$. The swimmer itself is modeled as a spherical squirmer of radius $R$ (SM-Fig. \ref{fgr: SMnum}). 

To quantify the degree of confinement, we define a dimensionless parameter:
\begin{equation}
w = \frac{h}{2R_d},
\end{equation}
which compares the channel half-height to the swimmer radius. Values of $w \sim \mathcal{O}(1)$ correspond to tight confinement, while $w \gg 1$ indicates effectively unbounded motion in the vertical direction.

The flow induced by the squirmer generates a pressure distribution on the walls, which is computed using the regularized Stokeslet method described previously. These hydrodynamic pressures serve as normal tractions acting on the elastic substrates. To compute the resulting deformation, we use an extension of the classical Boussinesq solution. 
For a vertical point force $f$ applied at position $\mathbf{x}'$ on the surface of an elastic half-space, the vertical displacement at a point $\mathbf{x}$ on the same surface is given by
\begin{equation}
\delta(\mathbf{x}) = \frac{(1 - \nu^2)}{\pi E} \cdot \frac{f}{r},
\end{equation}
where $\nu$ and $E$ are the Poisson's ratio and elastic coefficient of the elastic half-space, and $r$ denotes the Euclidean distance between source and observation points.
This result can be generalized to distributed tractions by superposing the contributions from all traction elements.

In our context, the hydrodynamic pressure $p(\mathbf{x})$ computed at the wall from the Stokes flow is interpreted as a normal traction distribution. We numerically integrate this pressure field over the surface of the wall to obtain the total vertical deformation at each wall point using:
\begin{equation}
\delta_z(\mathbf{x}') = \frac{1 - \nu^2}{\pi E} \int_{B} \frac{p(\mathbf{x})}{r}\, dS(\mathbf{x}),
\end{equation}
where $B$ denotes the wall surface domain and $\mathbf{x}$ and $\mathbf{x}'$ are coordinates on the wall. Since the elastic half-space model assumes an infinite substrate below the wall, this formulation accounts for long-range elastic effects and ensures mechanical consistency without requiring an explicit discretization of the wall interior.

This coupling of hydrodynamic pressure with substrate deformation through the Boussinesq integral provides a computationally efficient and physically realistic means to include substrate compliance. It allows us to study how wall elasticity alters the swimmer’s motion and how swimming-induced flows feed back onto the mechanical environment. In simulations, we compute $\delta_z(\mathbf{x}')$ at each time step and update the effective wall position, which in turn modifies the swimmer–boundary hydrodynamic interaction in subsequent steps.

\subsection*{2.3 Numerical simulation}
\subsubsection*{2.3.1 Method of regularized Stokeslet}
As mentioned earlier, the Stokeslet kernel is singular. 
To address this issue, the method of \emph{regularized Stokeslets} introduces a smooth approximation to the singular kernel by spreading the point force over a small region of size $\varepsilon > 0$. 
The resulting regularized kernel $S_{ij}^{\varepsilon}(\mathbf{x}, \mathbf{X})$ retains the essential far-field properties of the Stokeslet while removing the singularity at $\mathbf{x} = \mathbf{X}$, enabling stable numerical computation.

The fluid velocity at a point $\mathbf{x}$ is then approximated using the regularized boundary integral formulation:
\begin{equation}
u_i(\mathbf{x}) = -\frac{1}{8\pi} \int_{\partial D \cup B} S_{ij}^{\varepsilon}(\mathbf{x}, \mathbf{X}) f_j(\mathbf{X})\, dS(\mathbf{X}),
\end{equation}
where $S_{ij}^{\varepsilon}(\mathbf{x}, \mathbf{X})$ is the regularized Stokeslet kernel with smoothing parameter $\varepsilon$, $\mathbf{f}(\mathbf{X})$ is the unknown traction at point $\mathbf{X}$ on the surface, and the integration is performed over both the swimmer surface $\partial D$ and the no-slip boundary $B$ corresponding to the two walls of the channel. 
This regularized formulation forms the foundation for the nearest-neighbor method used to solve for the swimmer’s motion and the surrounding flow field.

To ensure that the swimmer remains force- and torque-free, we impose the following constraints:
\begin{align}
\int_{\partial D} \mathbf{f}(\mathbf{X})\, dS &= \mathbf{0}, \\
\int_{\partial D} (\mathbf{X} - \mathbf{x}_0) \times \mathbf{f}(\mathbf{X})\, dS &= \mathbf{0}.
\end{align}

\subsubsection*{2.3.2 Numerical implementation}
For numerical discretization, the swimmer’s surface and the boundary are each discretized by a set of points $\{\mathbf{x}[n]\}_{n=1}^{N}$, where the unknown traction vectors $\mathbf{f}[n]$ are defined, and a denser set of quadrature points $\{\mathbf{X}[q]\}_{q=1}^{Q}$ used for evaluating surface integrals. The quadrature weights associated with these points are denoted $w_q$.

The key feature of the nearest-neighbor regularized Stokeslet method is that the traction at quadrature points is approximated by assigning each quadrature point to its nearest force point. 
This gives rise to a nearest-neighbor matrix $\nu[q,n]$ defined as:
\[
\nu[q,n] = \begin{cases}
1 & \text{if } \mathbf{x}[n] \text{ is the nearest force point to } \mathbf{X}[q], \\
0 & \text{otherwise}.
\end{cases}
\]

Using this mapping, the discretized velocity at each force point becomes:
\begin{equation}
u_i(\mathbf{x}[m]) \approx -\frac{1}{8\pi} \sum_{n=1}^{N} f_j[n] \left( \sum_{q=1}^{Q} S_{ij}^{\varepsilon}(\mathbf{x}[m], \mathbf{X}[q]) \nu[q,n] w_q \right),
\end{equation}
for $m = 1, \dots, N$. 
This results in a system of $3N$ linear equations. 
To these, we append six additional equations for the force and torque balance constraints, leading to a total of $3N + 6$ unknowns (the $3N$ components of traction, and the 3 components each of $\mathbf{U}$ and $\boldsymbol{\Omega}$).

In the presence of a no-slip boundary, such as a wall or surface, the boundary is also discretized into force and quadrature points. 
On the boundary, the fluid velocity must be zero:
\begin{equation}
\mathbf{u}(\mathbf{x}) = \mathbf{0}, \quad \text{for } \mathbf{x} \in B.
\end{equation}
The same nearest-neighbor discretization is used for boundary points, and their contributions are included in the overall linear system. 
The unknown tractions on the boundary are solved simultaneously with those on the swimmer.

Once the linear system is assembled, it is solved for the unknown tractions, swimmer velocity $\mathbf{U}$, and rotation rate $\boldsymbol{\Omega}$. 
The swimmer's position and orientation are then updated over time by solving the following system of ordinary differential equations:
\begin{align}
\dot{\mathbf{x}}_0 &= \mathbf{U}, \\
\dot{\mathbf{b}}^{(i)} &= \boldsymbol{\Omega} \times \mathbf{b}^{(i)}, \quad i = 1, 2,
\end{align}
where $\{\mathbf{b}^{(1)}, \mathbf{b}^{(2)}, \mathbf{b}^{(3)} = \mathbf{b}^{(1)} \times \mathbf{b}^{(2)}\}$ define the swimmer’s body-fixed orthonormal frame. 

To compute the flow field at arbitrary points in the domain, the velocity is evaluated using the same regularized Stokeslet expression, now with known tractions:
\begin{equation}
u_i(\mathbf{x}) = -\frac{1}{8\pi} \sum_{n=1}^{N} f_j[n] \left( \sum_{q=1}^{Q} S_{ij}^{\varepsilon}(\mathbf{x}, \mathbf{X}[q]) \nu[q,n] w_q \right).
\end{equation}

In typical simulations, a squirmer surface is discretized using $N = 96$ force points, with $Q = 300–500$ quadrature points. 
Boundary walls are discretized similarly. 
The regularization parameter $\varepsilon$ is chosen based on the local grid on the squirmer surface spacing. 
The quadrature discretization is typically 3–5 times finer than the force discretization to ensure accuracy.

\section*{III. Additional data}
\subsection*{3.1 Experimental data on swimming dynamics in soft $(\mathcal{P} = 0.6)$ microchannel}
\begin{figure*}[ht]
\centering
\includegraphics[width=0.6\textwidth]{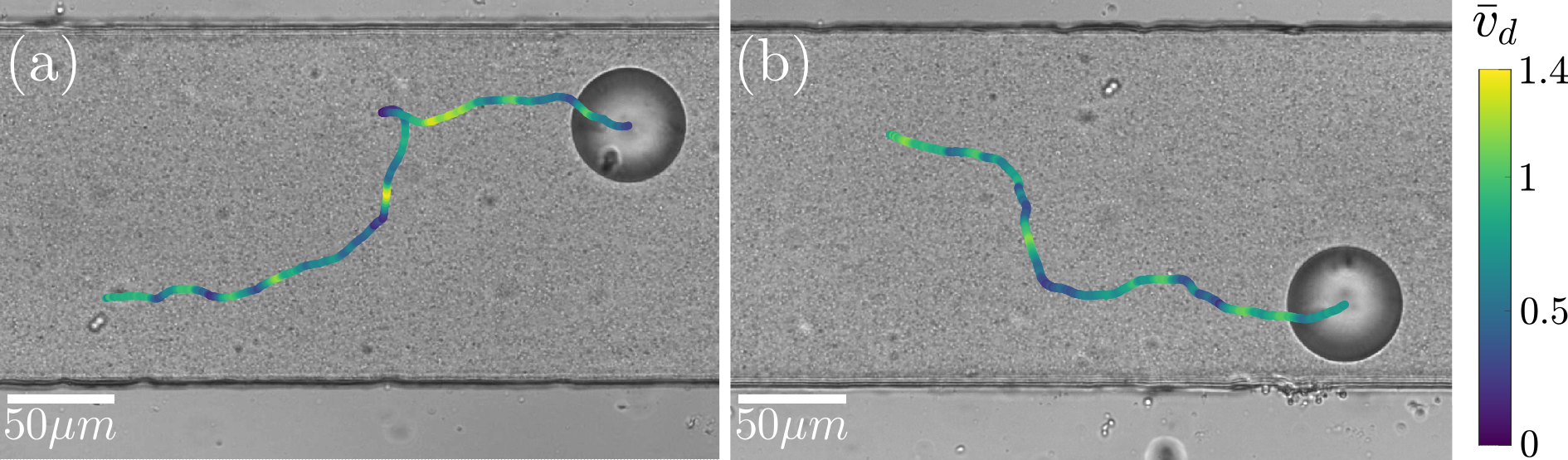}
 \caption{\label{fgr: SM4}\textbf{(a,b)} Swimming dynamics of active droplet in soft microchannel.}
\end{figure*}

Some additional experiments on the active droplets self-propelling in the soft microchannel ($\mathcal{P} = 0.6$) are included in SM-Figs. \ref{fgr: SM4}(a) and (b). The experiments are carried out by adding fluorescent tracer particles in the $7.5\%$ aqueous TTAB solution using bright field microscopy to ensure there is no effect of the background flow in the continuous medium, and the flow is ensured to be quiescent. The trajectory of the droplets is color coded with the instantaneous velocity, which is lower than the average velocity in rigid $(\mathcal{P} = 5)$ micro-channel (see Fig. 1 (i) in the main text).

\subsection*{3.2 Calculation of mean squared displacement (MSD)}

\begin{figure*}[ht]
\centering
 \includegraphics[width=0.5\textwidth]{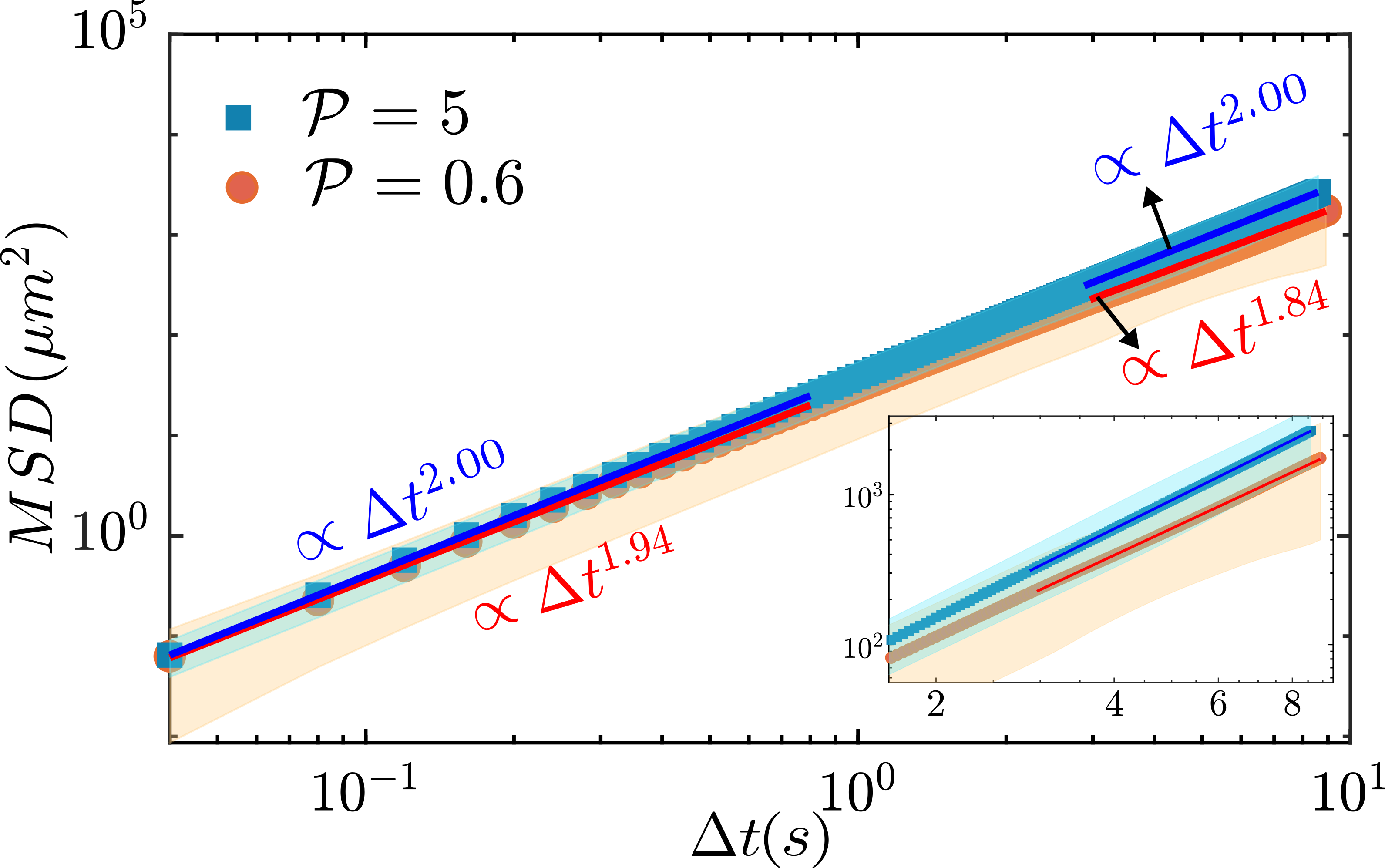}
 \caption{\label{fgr: SM6} Mean Squared Displacement (MSD) as a function of time for droplet swimming in rigid and soft microchannel. For rigid microchannel $(\mathcal{P} = 5)$, the slope of the log-log plot comes out to be 2, which illustrates ballistic behaviour over longer time scales.  For soft microchannel $(\mathcal{P} = 0.6)$, the slope is $> 1$, showing super-diffusive (directed motion) behavior, and it reduces over longer time scales.}
\end{figure*}

Mean squared displacement is calculated for each of the several datasets corresponding to active droplet motion in rigid ($\mathcal{P}=5$) and soft ($\mathcal{P}=0.6$) microchannels. 
For each time difference ($\Delta t$), MSD values of all datasets corresponding to respective microchannel rigidity are averaged. 
The figure (SM-Fig. \ref{fgr: SM6}) shows averaged MSD values plotted against $\Delta t$ for datasets corresponding to rigid and soft microchannels in a log-log scale, with standard deviations in MSD as error region. 
For small time scales $\Delta t$, we see that $MSD \propto \Delta t^2$ for active droplets in $\mathcal{P}=5$ microchannels and $MSD \propto \Delta t^{1.94}$ for active droplets in $\mathcal{P}=0.6$ microchannels. 
This shows that for smaller time scales, the active droplet motion in soft and rigid microchannels can be described as quasi-ballistic. 
For larger time scales $\Delta t$, active droplets still have $MSD \propto \Delta t^2$ (meaning that they still retain quasi-ballistic behavior) in rigid microchannels ($\mathcal{P}=5$), whereas active droplets in soft microchannels ($\mathcal{P}=0.6$) show $MSD \propto \Delta t^{1.84}$, which deviates from quasi-ballistic behavior. 

\subsection*{3.3 Experimental data on swimming dynamics in rigid $(\mathcal{P} = 5)$ and soft $(\mathcal{P} = 0.6)$ microchannel}

\begin{figure*}[ht]
\centering
 \includegraphics[width=0.55\textwidth]{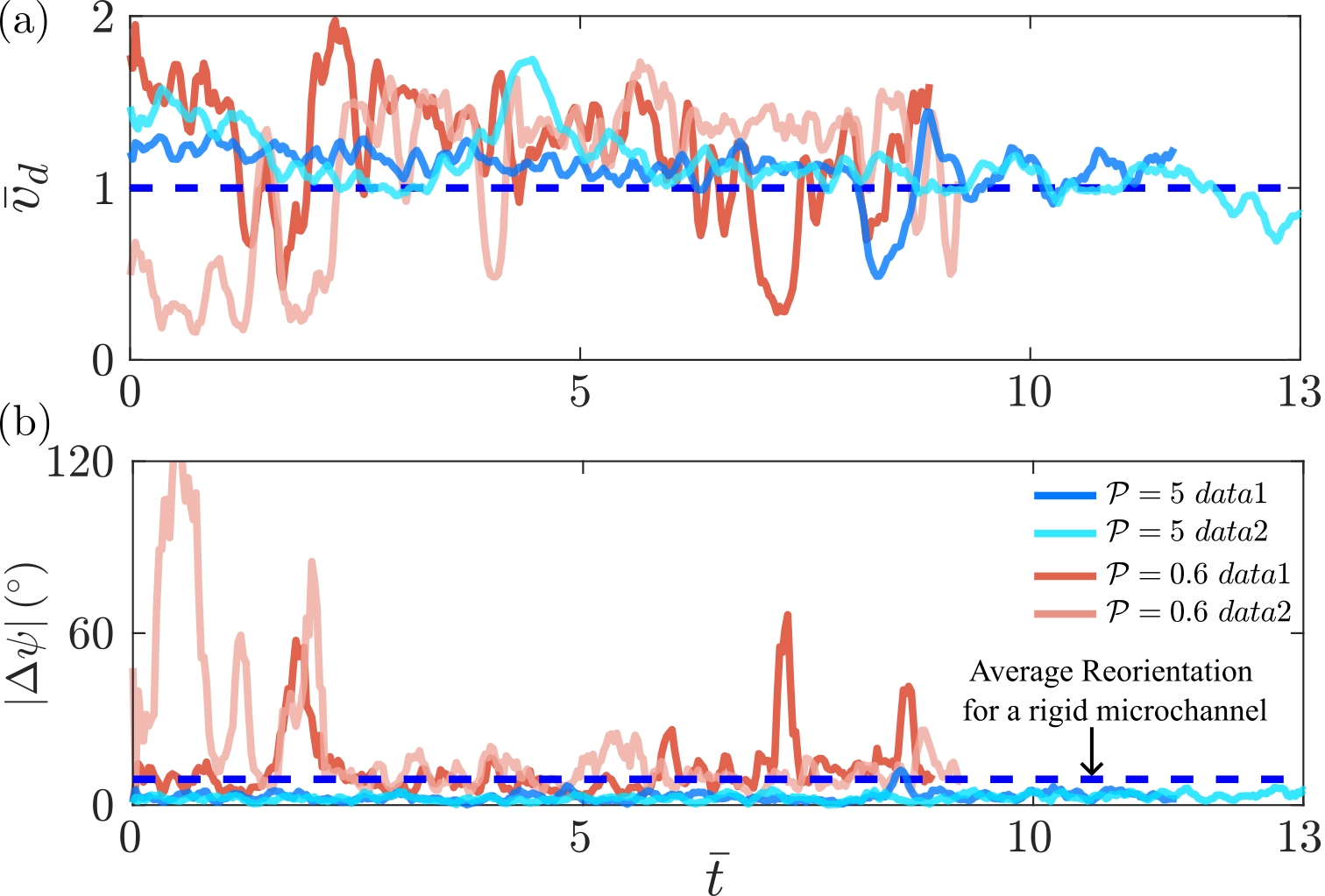}
 \caption{\label{fgr:SM5}Comparative data of instantaneous swimming velocity and change in orientation of active droplet in rigid and soft microchannel. \textbf{(a,b)} instantaneous swimming velocity of the droplet correlated with change in orientation over time.}
\end{figure*}

Additional data sets comparing the instantaneous swimming velocity and the change in orientation of the droplet in rigid $(\mathcal{P} = 5)$ and soft $(\mathcal{P} = 0.6)$ microchannel are presented in SM-Figs. \ref{fgr:SM5}(a) and (b). In the rigid microchannel, the droplet swims steadily with a higher average speed compared to the soft microchannel, as shown in SM-Fig. \ref{fgr:SM5}(a). The average change in orientation for the droplet in the rigid channel remains below $\approx$ 9$^\circ$, whereas in the soft microchannel, the droplet exhibits more frequent and larger reorientation events, as illustrated in SM-Fig. \ref{fgr:SM5}(b).

\subsection*{3.4 Dynamics of the active droplet swimming along an elasticity gradient}
\begin{figure*}[ht]
\centering
 \includegraphics[width=0.6\textwidth]{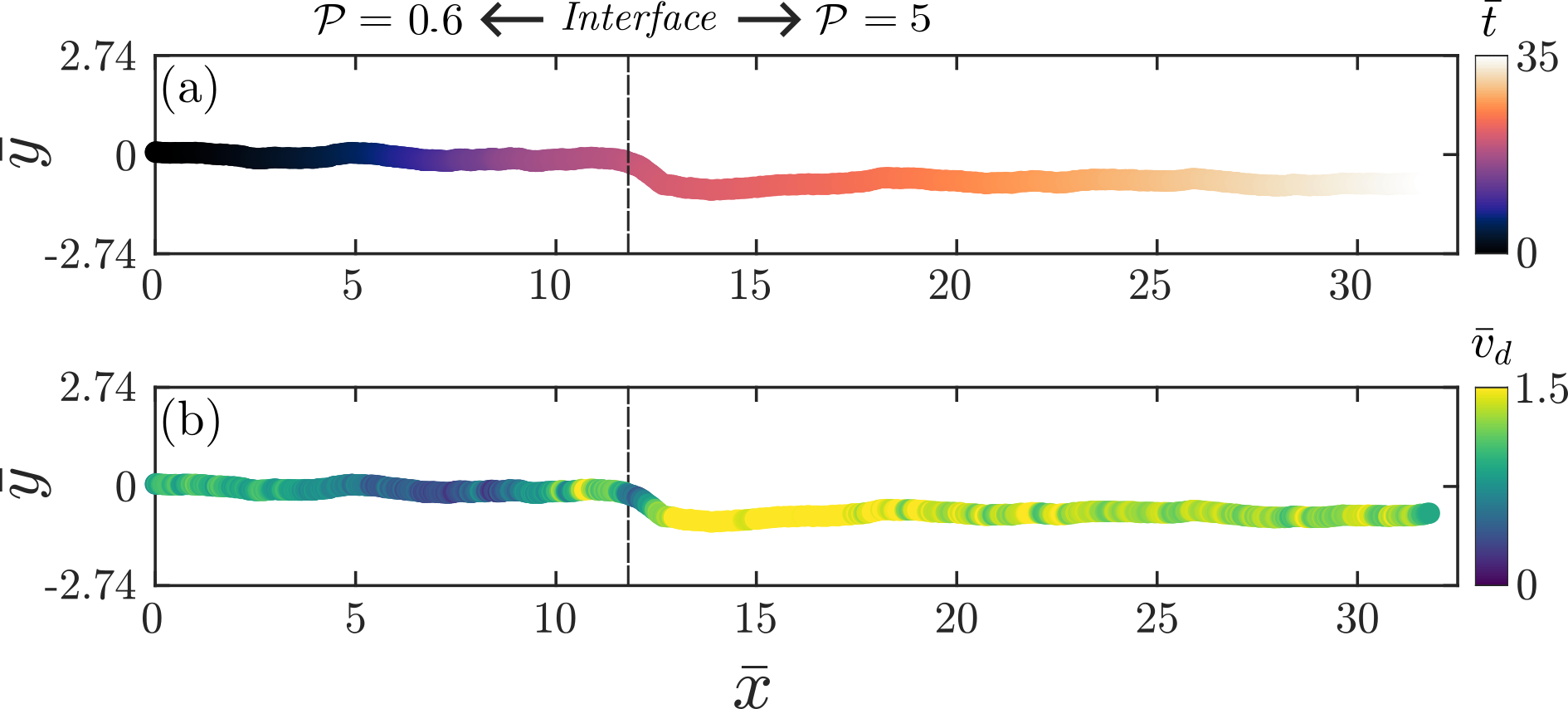}
 \caption{\label{fgr:SM7}Change in the dynamics of active droplet swimming along the increasing elasticity gradient ($\mathcal{P} =5 \ to \ \mathcal{P} =0.6$). \textbf{(a)} Droplet trajectory of the droplet color coded with the normalized time $\bar{t}$; \textbf{(b)} Droplet trajectory of the droplet color coded with the normalized velocity $\bar{v}_d$.}
\end{figure*}

\begin{figure*}[ht]
\centering
 \includegraphics[width=0.6\textwidth]{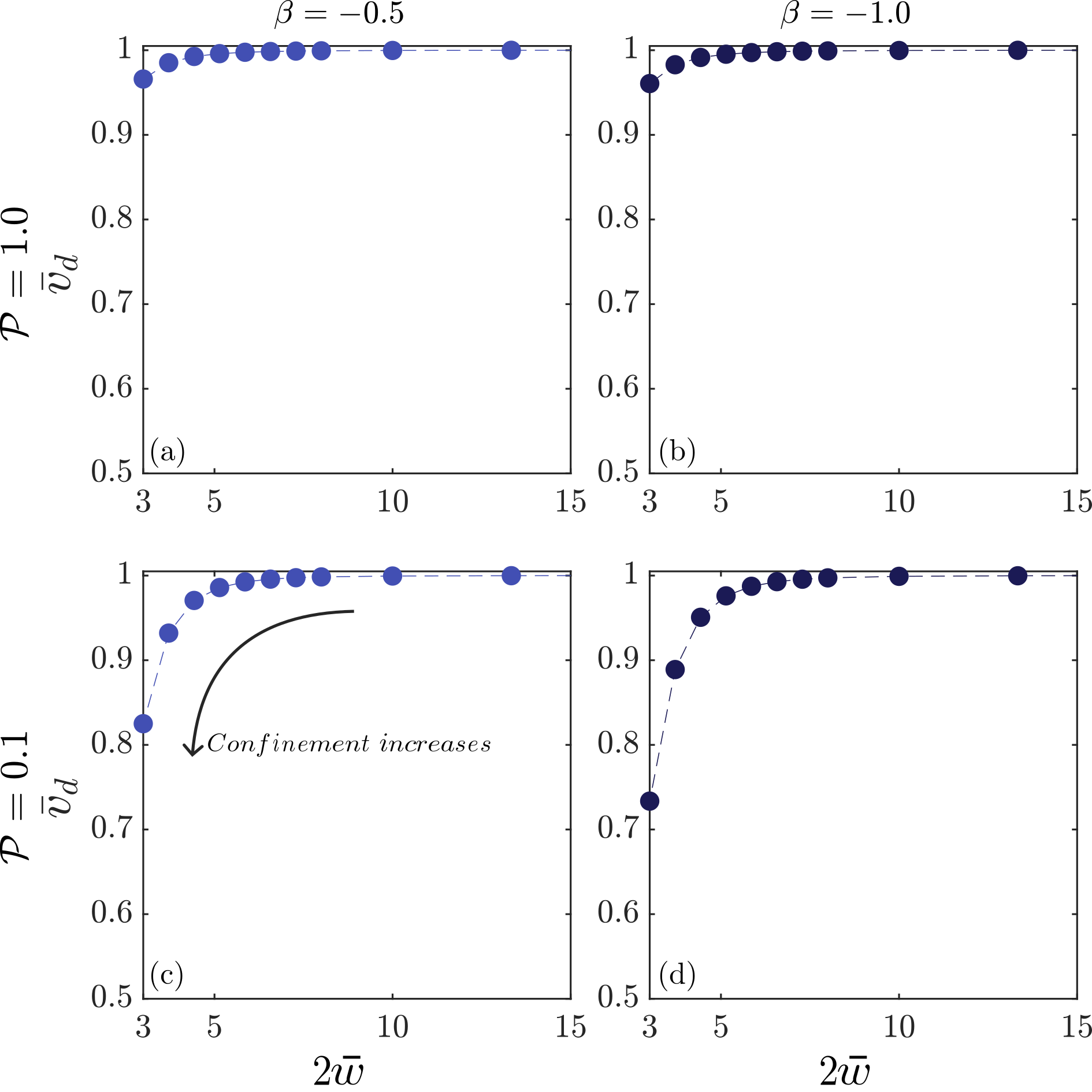}
 \caption{\label{fgr:SM8}{Numerical simulations depicting the effect of increasing the tightness of the microchannel on the normalized swimming velocity. \textbf{(a)} Variation of normalized velocity ($\bar{v}_d$) with confinement ($2\bar{w}$) for $\mathcal{P} = 1.0$, $\beta = -0.5$; (\textbf{b}) for $\mathcal{P} = 1.0$, $\beta = -1.0$; (\textbf{c}) for $\mathcal{P} = 0.1$, $\beta = -0.5$;  (\textbf{d}) for $\mathcal{P} = 0.1$, $\beta = -1.0$.}}
\end{figure*}

We prepare the compound microchannels having an increasing elasticity gradient, i.e., the Young's modulus ($E$) of the microchannel increases from $35 kPa$ to $380 kPa$. AS the droplet self-propels in the microchannel, it experiences a change in the mechanical properties of the confinement and adopt to its intrinsic behavior reported in the present work (SM-Fig. \ref{fgr:SM7}(a)). The droplet undergoes a run-and-tumble-like motion when it is swimming in the soft microchannel ($\mathcal{P} = 0.6$). The droplet swims with a lower instantaneous velocity in the soft region of the microchannel which is substantiated experimentally and numerically in this work (SM-Fig. \ref{fgr:SM7}(b)). Now as the droplet comes near the transition zone between the soft part and the rigid part (shown by dashed line in SM-Fig. \ref{fgr:SM7}) of the microchannel, the instantaneous swimming velocity starts increasing and the droplet undergoes changes its swimming orientation towards one of the confining side walls and swims along it (SM-Fig. \ref{fgr:SM7}(b)). The droplet recovered its intrinsic swimming behavior when it enters the rigid part of the compound microchannel.  
\newpage
\subsection*{3.5 Additional numerical simulation results}

\newpage
\section*{IV. Supplementary videos}
\begin{enumerate}
    \item  Swimming dynamics of the active droplet in a rigid microchannel ($\mathcal{P}=5$) and in a soft microchannel ($\mathcal{P}=0.6$) using bright field microscopy. The orientation of the velocity vector is shown using the red arrow at the centroid of the droplet; for video click (\href{https://drive.google.com/file/d/1x5iCuJL20AuIdBLnPZpIN1ylXclXZ9-a/view?usp=sharing}{\textbf{Video S1}}).
    
    \item Flow field generated by the active droplet in rigid microchannel ($\mathcal{P}=5$) and the changes in the flow field generated by the active droplet in soft microchannel ($\mathcal{P}=0.6$) visualized using the fluorescence microscopy ($\mu$PIV) experiments; for video click (\href{https://drive.google.com/file/d/1Xl3VeXTOIE63rDUwEn0vw5VlEW2zxAkY/view?usp=sharing}{\textbf{Video S2}}).
    
    \item Swimming dynamics of the active droplet swimming across the elasticity gradient from soft ($\mathcal{P}=0.6$) region to rigid ($\mathcal{P}=5$) in the same microchannel, the trajectory is color coded with the instantaneous velocity magnitude and vector represents the swimming orientation; for video click (\href{https://drive.google.com/file/d/1bFq-dRVIW0QxlBTKkHEq4KJXBBtBSwI_/view?usp=sharing}{\textbf{Video S3}}).
         
     \item Top: Fluorescence microscopy experiments in soft ($\mathcal{P}=0.6$) microchannel along with the corresponding velocity vector. Bottom: Intensity calculated at a distance of $\sim 22\mu m$ around the circumference of the active swimmer plotted over time; for video click 
     (\href{https://drive.google.com/file/d/1UBB1IP-Q3bfOx9dnyjcIoDh7u5-62Dcl/view?usp=sharing}{\textbf{Video S4}}).
     
\end{enumerate}


\end{document}